\documentclass[12pt,a4paper]{article}
\usepackage[T1]{fontenc}
\usepackage[utf8]{inputenc}
\usepackage[english]{babel}
\usepackage{amsmath,amssymb}
\usepackage{graphicx}
\usepackage[section]{placeins}
\usepackage{geometry}
\usepackage{authblk}
\usepackage[super,sort&compress,comma]{natbib}
\usepackage{hyperref}
\hypersetup{hidelinks}
\newcommand{\balance}{}

\title{Bottom-Up Surface-Enhanced Raman Scattering for the Detection of Micro-Objects and Globular Analytes}
\author[1]{Andrey K. Sarychev\thanks{Corresponding author: sarychev\_andrey@yahoo.com; phone: +79264714000}}
\author[2,3]{Milena S. Shestopalova}
\author[2,3,4]{Denis S. Korzhov}
\author[1]{Andrey V. Ivanov}
\author[1]{Igor V. Bykov}
\author[1]{Konstantin N. Afanasev}
\author[4,5]{Nikita N. Osipov}
\author[4]{Dmitry V. Basmanov}
\author[2,3]{Vladimir A. Oleinikov}
\author[2,6]{Konstantin E. Mochalov}
\affil[1]{Institute for Theoretical and Applied Electromagnetics of the Russian Academy of Sciences, Moscow 125412, Russia}
\affil[2]{Shemyakin-Ovchinnikov Institute of Bioorganic Chemistry of the Russian Academy of Sciences, Moscow 117997, Russia}
\affil[3]{National Research Nuclear University ``MEPhI'' (Moscow Engineering Physics Institute), Moscow 115409, Russia}
\affil[4]{Lopukhin Federal Research and Clinical Center of Physical-Chemical Medicine of Federal Medical Biological Agency, Moscow 119435, Russia}
\affil[5]{Institute of Microelectronics Technology and High Purity Materials of the Russian Academy of Sciences, Chernogolovka, Moscow Region 142432, Russia}
\affil[6]{RUDN University, Moscow 117198, Russia}
\date{}
\begin{document}
\maketitle
\begin{abstract}
Surface-enhanced Raman scattering (SERS) is typically achieved
by depositing the analyte on a metal or semiconductor substrate. We propose
a bottom-up approach, enabling SERS by depositing a metal on the surface of
the object being studied. A Mie-theory analysis of electric-field enhancement
in dielectric core--shell particles is presented. A correlative approach
combining SERS, atomic force microscopy, and scanning electron microscopy
is employed to model the conditions required for sensitive SERS detection
and spectral analysis of polystyrene microspheres (PMSs) as models of globular
and vesicle-like objects. Based on these findings, we propose a SERS-active
system that provides strong signal enhancement (up to four orders of magnitude
for a single globular object) while maintaining spectral fidelity, where PMSs
act both as model analytes and as integral components of a metal--dielectric
resonator.
\end{abstract}

\section{Introduction}
Raman scattering (RS) is an effective tool for detecting 
and studying not only molecules but also artificial and 
natural nanoobjects, such as nanoplastics, virions, exosomes, 
and protein globules.
Surface-enhanced Raman scattering (SERS) enables spectroscopic 
data to be obtained from individual nano- and microobjects.
The advancement of technological capabilities 
for fabricating plasmonic 
resonator nanostructures largely determines the growing interest in 
surface-enhanced Raman spectroscopy (SERS), 
both from the perspective 
of fundamental studies and practical applications
\cite{ZouMaZheng2025, 
PerezJimenezLyuLu2020,RanasingheWangHuang2024,%
HanRodriguezHaynes2021,PilotSignoriniDurante2019}. 
Plasmonic 
resonator nanostructures are most often based on noble metal surfaces, 
realized as clustered arrays of nanoparticles 
\cite{GresillonAigouyBoccara1999,LiuLiuTang2019} or ordered 
morphologies such as corrugated {metal} films, sharp tips, nanopores, or 
nanodisks 
\cite{KukushkinAmbartsumyanSubekin2023,%
LagarkovBoginskayaBykov2017,IvanovBykovBarbillon2024,%
BarbillonIvanovSarychev2019}. 
Such architectures allow resonance frequencies to 
be tuned to the Stokes shifts of analytes, while simultaneously improving 
the stability and reproducibility of SERS spectra. 
 {In SERS-based analyte detection, 
	the pure Raman spectrum of the analyte 
	should be determined for the particular} 
substrate. This restriction is not essential for the SERS investigation 
of {microplastics or biological objects such as exosomes and virions.}
 {For microscopic objects, the Raman signal depends on}
the structure of a SERS substrate.
Sometimes this feature of SERS is {considered a} limitation for
its widespread adoption as an analytical 
technique \cite{AverkievRodriguezFatkullin2024,%
PerezJimenezLyuLu2020,PangYangHe2016}. 
Considerable effort has therefore been devoted to 
understanding {the structure} of SERS spectra that depend on
chemical or electromagnetic interactions between 
 {micro-objects} and plasmonic nanostructures 
\cite{HanRodriguezHaynes2021,HanJiZhao2017, 
VenkatachalamBoerioCarnevale1988}. 

Dielectric resonant nanostructures have been 
investigated as ``plasmon-free'' SERS platforms, eliminating ohmic losses 
in metals and reducing analyte overheating
 \cite{AlessandriCarlettiFerroni2020,GuasoniBontempiNeshev2016,%
AlessandriLombardi2016}. 
These systems offer enhanced spectral stability 
and reproducibility
\cite{BontempiCarlettiDeAngelis2016,
BontempiVassaliniAlessandri2018}.
Yet their sensitivity is 
significantly lower than {that of metal} nanostructures. This trade-off has 
motivated the development of hybrid strategies that combine dielectric 
 {metasurfaces} with thin {metal} coatings, enabling the creation of open 
plasmonic resonators with improved performance. 
Experiments have shown the versatility of 
metal–dielectric metasurfaces for Raman enhancement
 \cite{Barbillon2022,SarychevBykovBoginskaya2020,%
 SarychevLagarkovIvanov2017}. 

Within this broader field, {polystyrene microspheres (PMSs) have} 
become one of the most versatile dielectric building blocks. Their 
monodispersity, tunable size, 
and ability to self-assemble make them ideal 
for nanosphere lithography, a scalable route 
to ordered plasmonic arrays 
\cite{HaynesVanDuyne2001}. 
Depositing gold films on PMS arrays produces 
highly effective SERS 
substrates across the visible–NIR range \cite{BaiaBaiaPopp2006}, 
while silver 
and gold nanoshells on PMS cores exhibit tunable plasmonic 
resonances dependent on shell thickness 
\cite{YongSahooSwihart2006}. 
More complex {morphologies, such as 
silver-coated} inverse opals formed from PMS 
templates \cite{MuHwangChang2011}
and nanoparticle-decorated beads
\cite{MirSimonMorlaFolchGisbertQuilis2015} 
further enhance the density of plasmonic "hot spots.'' 
Advances in synthesis, including chemical reduction routes for
 PMS/Ag composites \cite{ZhuWuYan2013}, 
 uniform nanoparticle deposition 
\cite{JunfanTaoJie2020}, and hybrid nanocap architectures 
\cite{ShiGaoShang2023}, have enabled 
integration into microfluidic chips 
\cite{PengXuYi2024} and anisotropic conductive films 
\cite{HaoChenChen2025}. 
Together, these approaches illustrate how 
PMS-based substrates provide a low-cost yet powerful 
foundation for SERS device engineering. 

Beyond plasmonic {coatings,} the optically transparent 
microspheres have been exploited for near-field imaging
beyond the diffraction limit
 \cite{WangGuoLi2011,VergelesSarychevTartakovsky2017}.
Individual microspheres can focus incident light into photonic nanojets, 
leading to significant Raman signal amplification 
\cite{KasimTingMeng2008,DuKasimYou2011}. 
Self-assembled arrays of PMS microlenses {placed on 
silicone, gold, or silver substrates with deposited analytes
 enhance the RS signal by almost two orders of 
 magnitude}\cite{YanXingJia2015,XingYanFeng2017,
HongZhouXu2020,Ogura2022,WangYanMi2022}.
For example, when gold nanoparticles containing the analyte
are coated with a densely packed array of dielectric microspheres with a 
diameter $ \sim 5 \, \mu m ${, the RS is enhanced by two orders of 
	magnitude}\cite{WangYanMi2022}.
In a sense, this is a "bottom-up" approach,
where the amplifiers coat
the material being studied.
 {Hybrid designs} that {combine} dielectric focusing with plasmonic 
coatings — such as metal-coated micro-objects — 
offer synergistic effects, providing both high enhancement 
factors and improved reproducibility 
\cite{WangHaoYin2019,SarychevIvanovLagarkov2019}.

The optical versatility of PMS extends well beyond Raman 
spectroscopy. Microsphere cavities fabricated by chemical vapor 
deposition have been used to tune excitonic properties in 2D 
semiconductors such as MoS$_2$ \cite{MiZhangZhao2017}, while 
monolithic microsphere–WSe$_2$ cavities have enabled low-threshold 
lasing \cite{YuXingDai2022}. Laser-irradiated PMS can even 
function as multimode light sources \cite{LiuTieChen2022}, and 
rare-earth-doped spherical cavities demonstrate their potential in 
microlasers \cite{OuyangKangZhang2019} 
(see, however, discussion in 
 \cite{StefancuAizpuruaAlessandri2024,
 LopezCarrascoGuimaraes2025}). 
In parallel, PMS-based SERS 
probes have been developed for biomedical imaging, with encapsulated 
SERS tags offering high sensitivity and long-term stability 
\cite{YuWangMei2019}. Environmental applications include the 
detection of polystyrene microplastics, where SERS provides molecular 
fingerprints for {pollutant identification} \cite{MikacRigoHimics2023}. 
On the materials side, Raman studies of PMS 
itself have revealed information on its 
glass transition\cite{MenezesReyerMarletta2017}. 

These advances underline the relevance of dielectric {microspheres} in two 
high-impact areas: biomedicine and environmental monitoring. 
In diagnostics, SERS has been applied to extracellular vesicles 
and exosomes{,} 
\cite{GuerriniGarciaRicoLoghlen2021,ChenQiu2025,%
LeeCarneyHazari2015}
including recent detection of a single exosome
\cite{IvanovMochalovKorzhov2025,%
MochalovKorzhovShestopalova2025,%
ShestopalovaKorzhovAfanasyev2026}
as well as to virus detection 
\cite{GarsuaultElMessaoudiPrabakaran2023,LvZhangZhao2023,%
SarychevSukhanovaIvanov2022}. 
In environmental science, Raman-based methods have been used 
for assessing nanoplastic contamination 
\cite{SchroeterVentura2022,TrevisanRanasingheJayasundara2022,
MoghaShin2023}. 
Notably, these analytes, i.e., extracellular vesicles, viral particles, 
and {nanoplastics,} are often globular or vesicle-like 
in morphology, making them conceptually similar 
to dielectric microspheres. 
This methodological parallel further motivates the use 
of PMS as model resonators for the design 
and optimization of {SERS methods}. 

In this paper{,} we consider Raman spectra obtained from single 
 {compact microscopic objects, such as microplastics, virions, and exosomes, 
that are covered by a thin metal shell to obtain} their SERS {signatures}. 
 {A generic analytical theory} is developed for the EM field 
in the dielectric {core--shell resonator}. 
The theory is based on the Mie expansion
 \cite{Mie1908,AdenKerker1951} and our previous works
 \cite{SarychevIvanovBykov2023DD,SarychevIvanovBykov2023QE,
 SarychevIvanovBykov2024QE,SarychevIvanovBykov2024ModEl}. 
 {The investigation} of the electrodynamics of {core--shell resonators 
 has a long history. Quasistatic theory} 
 {(\cite{LandauLifshitzPitaevskii1993}, Sec.\,8; 
 \cite{NeevesBirnboim1989})}
predicts that hollow metal nanoparticles could substantially shift
plasmon resonances to longer wavelengths than 
solid metal nanoparticles.
The resonant frequency is controlled 
by the spherical shell dimensions.
 {Dielectric-core/metal-shell nanoparticles, known as nanoshells, 
 were first realized in Au$_2$S--Au and then SiO$_2$--Au materials}
\cite{ZhouHonmaKomiyama1994,AverittSarkarHalas1997}. 
Varying the {ratio between the inner and outer radii} of the gold
shell tunes the plasmon resonance, and therefore the wavelength of light
absorbed by the particle \cite{ProdanRadloffHalas2003}. 
Tuning the plasmon resonance beyond the visible and into 
the near-infrared region of the spectrum, a region of high {human-tissue}
 transparency, opened the door to a wide range of applications 
in nanomedicine\cite{Weissleder2001}.
Near-infrared {nanomedicine} spans diagnostics, such as {probe-based}
 imaging of tumor margins, as well as therapeutics, such as remotely
 triggerable drug or gene delivery. One of the most promising 
therapeutic modalities is plasmon photothermal cancer therapy
 \cite{RastinehadAnastosWajswol2019}. 
In this therapy, core--shell nanoparticles accumulate in tumor tissue 
via leaky tumor vasculature, and then are irradiated with a 
near-infrared laser. The tumor undergoes photothermal heating, 
resulting in selective hyperthermic cell death, without heating 
the adjacent nontumorous tissue. {Another example is a photothermal
 nanomedicine core--shell brain-stimulation system 
 that is proposed to reverse Parkinson’s disease} \cite{WuCuiBao2025}.
The core--shell plasmon particles restore degenerated dopamine 
neurons in {the brain} after in situ photothermal heating. 
Since {the frequencies} of Raman lines {shift slightly with increasing
temperature,} a core--shell particle can be used not only as 
 {a nanoheater but also as an optical thermometer} 
(see, e.g., \cite{SkinnerSalaSokolowski2025}).
 {A similar effect} is used in \cite{MenezesReyerMarletta2017}
for exploring the glass transition in the polymer polystyrene core.

Core--shell nanostructures can be obtained by metal deposition 
on {dielectric spherical particles}. The {rough} silver surfaces were
obtained by vapor deposition of Ag on {Teflon} spheres 
of uniform size and shape on glass substrates
 \cite{GoudonnetBegunArakawa1982}.
 Thick evaporated Ag films over a monolayer of polymer nanospheres
 {were studied by the Van Duyne group}. Deposition of a metal film over
 nanoparticles, known as FON, results in {effective} SERS, 
 {anti-Stokes-detected SERS}, and {surface-enhanced} hyper-Raman
 spectroscopy (SEHRS) 
\cite{VanDuyneHulteenTreichel1993,JacksonWestcottHirsch2003}.
Many features of the plasmon {resonances} in core--shell structures can 
be {understood using} plasmon hybridization theory
\cite{ProdanRadloffHalas2003,GradyHalasNordlander2004}.
In analogy to the hybridization of atomic orbitals 
in {molecular orbital theory,} the plasmon modes of coupled
 nanoparticles can be seen as {``hybridized''} plasmon modes resulting 
 from the interaction of the plasmons of the involved nanoparticles. 
 These interactions result in {low-energy
``bonding'' and high-energy ``antibonding''} plasmon modes,
 which {correspond} to symmetric and antisymmetric coupling of the modes. 
The effect of hybridization is most important when {a} core--shell 
resonator has no central symmetry 
\cite{CharnayLeeMan2003,KnightHalas2008,
JahnPatzeHidi2016,ZhuZhao2016,SouriHadilouNavid2021,
HadilouSouriNavid2022,AlsawaftaAbuzir2025}. 
For example, the symmetry of spherical core--shell nanoparticles is
destroyed by displacing the dielectric core from the center. 
The broken symmetry allows the hybridization of multipolar sphere 
and cavity modes. This leads to a couple of excitable 
plasmon resonances, which can be tuned by varying the offset 
of the core. The maximum field enhancement 
of these {structures, called nanoeggs, is located 
near} the thinnest part of the shell\cite{KnightHalas2008}.
 {As the core offset increases, the plasmon resonance 
corresponding to the maximum field enhancement 
becomes red-shifted until the core passes through the shell surface,}
ending up with a nanocup structure. 
Asymmetric Au-core -- Ag-shell nanorods with strong plasmon 
resonance are used as local second-harmonic generators
\cite{DingYangDeshpande2018}.

Some plasmon resonances in {core--shell structures} are ``dark'' since 
they {cannot be excited
by ordinary incident light, i.e., a plane wave}. 
Excitation of a dark localized plasmon in asymmetrically 
coated core--shell nanoparticles is theoretically possible if the incident 
plane wave is replaced by an optical {vortex}
\cite{TanakaHarajiriFujita2024}.
Asymmetric core--shell particles can be {designed} to give strong 
scattering in the optical frequency range by suitably tailoring the
 core--shell hybridization \cite{VarytisBusch2020}.
The electric field in {a} core--shell plasmon resonance is rather nonuniform
even for {a symmetric} core--shell resonator due to the various
 plasmons excited in the metal envelope
\cite{MannNascimentoDuplatKeul2017,LinLiaoChen2011}. 
Therefore, the local SERS is spatially nonuniform. 
The distribution of the local SERS $ G $ from {a} silver FON,
which is covered by {a} monolayer of benzenethiolate, {was investigated} in
\cite{FangSeongDlott2008}. The local SERS distribution 
$ P(G)\,dG $ was found to be a power law proportional to 
$ G^{-1.75} $, with minimum and maximum 
values of $ 2.8 \cdot 10^4$ 
and $4.1 \cdot 10^{10}$, respectively. 
It is interesting that some Raman peaks of benzenethiolate around 
$1500\,cm^{-1}$ increased by many orders {of} magnitude when {the FON 
was irradiated with a pulsed laser whose power exceeded a certain critical value}. 
We speculate {that} the effect could {result} from {the} photoinduced formation of
 {a metastable metal--molecule state that gives resonant SERS} 
\cite{DemuthSanda1981,MortonJensen2009,
ValleyGreeneltchVanDuyne2013,HernandezReynosoFainstein2023,
RanasingheSandersWang2026}.
 {In ref.}\,\cite{TognalliCortesHernandezNieves2011}{,}
SERS generated by gold nanoparticles covered by 
 {a few atomic layers of silver was investigated}. 
Even {a single} silver layer gives almost an order 
 {of} magnitude enhancement
of the Raman signal. {The authors} attribute this surprising result
to the rearrangement of the electronic charge density related 
to the presence of the Ag monolayer in the Au/Ag/molecule
(4-mercaptopyridine) complex. 
 {A core--shell SERS study showed} 
\cite{KurouskiPostiglioneDeckertGaudig2013}
 {that differences between the Raman lines of native insulin 
and insulin fibrils are determined by the size of the attached amino acid.}

 {SERS from a FON consisting of a $200\,nm$ 
 	silver film covering silica and
polystyrene microspheres was investigated} in the seminal paper 
\cite{GreeneltchBlaberHenry2013}.
It is shown that, by varying the size of the microspheres, 
the surface plasmon resonance can be tuned between 330 and 1840 nm.
Notably, there are predicted optimal microsphere sizes for 
each of the commonly used SERS laser wavelengths of 
$532, 633, 785 $, and $1064\,nm$. 
The Raman signal {increases substantially} when the silver film
is {activated} by deposition of the {``immobilized nanorod assemblies''} 
on its surface.
Using high-resolution SEM,
the activated film surface was shown to consist of an
array of randomly organized nanopillars.
 {A similar} silver FON was modified by a polycrystalline metal–organic
framework and used to measure SERS from solvents like benzene, 
toluene, {and} nitrobenzene{, which} weakly adsorb on {the} silver surface and {do not
produce a} Raman signal at a smooth metal surface
\cite{KrenoGreeneltchFarha2014}.
The effect of {FON activation was observed in the studies} 
\cite{WangWuXiao2014,GuoXiaoZhang2015}
where SERS {from a} gold FON ($ \sim 200 \, nm$ 
gold film deposited on closely packed PMS) was {substantially} 
 {boosted} by {sputtering} silver nanoparticles (diameter $40 \div 60 \, nm$) 
on the gold surface. 

SERS generated by {a 
FON made from closely packed polystyrene spheres 
covered by a $120\,nm$ silver film was investigated} in
\cite{LinLiaoChen2011}.
The maximum SERS is achieved when the core diameter is
about one {micron, with} the silver shell thickness fixed at 120 nm. 
Yet, the variation of the Raman enhancement is rather moderate: the 
enhancement factor $G$ changes 
{by a factor of \textit{seven}} when the core diameter 
changes from 0.8 to 1.6 $\mu m$. {More generally, the electromagnetic SERS
enhancement contains the local-field factors at both the laser 
 frequency $\omega$ and Stokes frequency $\omega_{S}$,}
\[
G(\mathbf r) \simeq
\left|\frac{E_{\rm loc}(\mathbf r,\omega)}
{E_{\rm in}(\omega)}\right|^2
\left|\frac{E_{\rm loc}(\mathbf r,\omega_S)}
{E_{\rm in}(\omega_S)}\right|^2
 \simeq \left|E_{\rm loc}/E_{\rm in}\right|^4,
\]
{when the two field-enhancement factors are approximately equal}
(see, e.g., \cite{BrouersBlacherLagarkov1997}), therefore,
the field enhancement could be effectively {damped} in the core--shell 
plasmon resonances in the case of the moderate values of $G$. 

In this work, we present {a generic analytical theory} for the electric 
field enhancement in dielectric {core--shell particles}.
The local field averaged over the internal surface 
of the metal shell is found. {That is,} we find the field 
at the {metal--dielectric} interface{, making it possible} to estimate the
 enhancement of RS and other {spectroscopic signals} from 
 an object surrounded by the metal envelope. 
 {Similarly,} we derive an equation for the field enhancement 
at the external surface of the metal shell. Therefore, SERS from
the molecules deposited on the surface of the metal shell
is estimated as {a} function of the diameter $2 a $ and thickness $ d $ 
of the metal envelope. We speculate that plasmon resonances 
correspond to the excitation {of plasmons} circulating in the thin
metal shell. 
 SERS is experimentally explored in
globular metal–dielectric resonant metastructures based on {PMSs
 employed as model analytes} mimicking globular 
proteins, vesicles, viruses, and related biological 
or environmental objects. For the first time{,}
RS from single PMSs {was collected}.
The morphological characteristics of individual {PMSs}, 
including those coated with silver nanoshells, 
were examined using atomic 
force microscopy (AFM) and scanning electron microscopy (SEM).
 We further investigated their optical response 
and SERS signal behavior as a 
function of excitation power. The results reveal the dual influence of 
the laser power{,} which simultaneously {enhances} Raman 
signals and {introduces spectral modifications}. Based on these findings, we 
provide practical guidelines for the design of metal–dielectric 
metastructures tailored to the SERS analysis of globular and vesicle-like 
analytes, emphasizing the balance between spectral fidelity 
and detection sensitivity.
\section{Experimental and Theoretical Methods}
 \label{Sec Surface-enhanced Raman scattering 
 from polystyrene microspheres (PMS)}
\subsection{Synthesis of PMS and fabrication of silver substrates}
\label{SubSec Synthesis of PMS and fabrication of silver substrates}
 
PMSs with an average diameter of $\sim 300\,nm $
and a refractive index of 
$ n_{1} =1.6$\,\cite{MarxMulholland1983}
were used in this 
study. The synthesis of polystyrene {PMSs} was performed by dispersion 
radical polymerization of styrene, following the procedures described in 
\cite{ChoShinHan2016}.
Cover glass substrates (Menzel Gläser Type I, $18 \times 18 \, mm$; 
Thermo Fisher Scientific, Waltham, MA) were used 
for the fabrication of metal--dielectric microresonators. 
The first step consisted of 
depositing a base silver layer onto the surface of the cover glasses. 
Prior to 
silver deposition, the substrates were thoroughly cleaned using Piranha 
solution (H$_2$SO$_4$/H$_2$O$_2$, 3:1), 
followed by repeated rinsing 
with bidistilled water and drying under a stream of compressed air. 
A $200\,nm$ thick base silver layer was then deposited onto the cleaned 
glass substrates. 
Metal deposition was carried out using a vacuum system equipped 
with an electron-beam evaporator. The substrates were fixed on 
a rotating turret, and the chamber was 
evacuated to a pressure of $1\,Pa$. 
A glow discharge was maintained for $30\,min$ at $1500\,V $ and 
$80\,mA$ to ensure surface conditioning. Subsequently, the chamber 
was pumped down to 2 $\times$ 10$^{-3}$\,Pa, and silver was 
evaporated from a molybdenum crucible placed in a water-cooled 
copper holder of the electron-beam evaporator. 
The accelerating voltage of the evaporator was $8\,kV$ with a beam 
emission current of $40 \div 50\,mA$. During deposition, a preliminary 
estimation of the film thickness was performed by monitoring 
the optical density of a reference sample. 
After deposition, the silver film thickness was determined 
by microinterferometry and was found to be approximately $200\,nm$. 
\subsubsection{PMS deposition on silver substrates 
	and {formation of} core--shell resonators}
	\label{SubSec PMS deposition on silver substrates 
and formation core--shell resonators}
Immediately after the deposition of the silver layer, 
 {PMSs} were deposited. For this purpose, the colloidal stock 
solution of the microspheres in Milli-Q water (Millipore) 
at {a} concentration of 
5\% was purified from residual monomers. 
Specifically, $2~mL$ of the PMS stock solution was centrifuged at 
$10\,000~rpm$ for $7~min$, after which the supernatant was 
removed and bidistilled water was added 
to restore the initial volume. 
This procedure was repeated twice. For the deposition procedure, 
the purified PMS stock solution was diluted twenty-fold with Milli-Q 
water (Millipore). The resulting suspension was sonicated in an 
ultrasonic bath for 6~min at room temperature. 

Subsequently, the metallized glass substrates were immersed {in} 
the diluted colloidal PMS solution at an angle of $ 20^{\circ} $ 
relative to the surface normal and then slowly withdrawn from 
the solution at a rate of $ 1~mm/min$. This process, accompanied 
by the formation of a positive meniscus, enabled the deposition 
of a sufficiently uniform, sparse layer of {PMSs} on the silver surface. 
The prepared samples were dried in a desiccator for at least $ 3~h $. 

The first characterization cycle of the obtained samples was
 performed using {AFM, SEM, and optical confocal
 microspectroscopy}. {The final step, performed after
 the spectro-topographic investigations, consisted of depositing a $20 \, nm$ 
 top silver layer} onto the surface according 
 to the procedure described above. Thus{,} core--shell metal-dielectric 
 {resonators were formed, with the investigated object serving as the core}.
 {An} aqueous stock solution of $0.5\, \mu M$ L-phenylalanine
 (Sigma Chemical Co.) was prepared to verify the intrinsic 
 SERS activity of the smooth silver surface.
\subsection{Scanning electron microscopy (SEM) measurements}
\label{SubSec Scanning electron microscopy (SEM) measurements}
SEM images of PMS metal--dielectric resonators were 
acquired using a Carl Zeiss Auriga Dual-Beam workstation operated in 
normal high-vacuum mode. An Everhart--Thornley secondary electron 
detector was employed with a $200\,p\mathrm{A}$ 
electron beam at $20~k \mathrm{V} $ landing 
energy and a working distance of $ 5.6\,mm $. 
\subsection{Combined {atomic} force microscopy (AFM) 
 and microspectroscopy measurements}
 \label{SubSec Combined Atomic force microscopy (AFM) 
 and microspectroscopy measurements}
PMS-based metal--dielectric resonators were investigated 
using {a} home-built setup, the ``System for probe--optical 
3D correlative microscopy'' (IBCh RAS,
 \url{http://ckp-rf.ru/usu/486825/}), which integrates confocal
 microspectroscopy with 2D/3D scanning probe microscopy 
 in a single apparatus~\cite{MochalovChistyakovSolovyeva2017}. 
Raman and SERS spectra were obtained using the upright confocal 
optical module of this setup, which includes an Andor Shamrock 
750 monochromator equipped with an Andor DU971P-BV CCD 
(Andor Technology Ltd.), two Ultrasteep long-pass edge filters
 (RazorEdge\textsuperscript{\textregistered}, Semrock) 
 {to suppress} the excitation laser, and a 
 Leica $\times$100/NA~0.75 objective lens. 
 A water-immersion Leica $\times$63/NA~0.9 objective lens 
 was used to test the intrinsic SERS activity of flat silver surfaces. 
 The excitation laser wavelength was $532\,nm$. 
 All spectra were 
 recorded with an exposure time of 1~s and 25 accumulations.
 The laser power was adjusted depending 
 on the experimental conditions to ensure that the Raman 
 band intensity remained within the CCD detection range. 
 The linear dependence of Raman intensity on laser power was 
 verified in all cases. Raman and SERS spectra were collected 
 in the spectral range of {$500-1800~cm^{-1}$}. 

AFM images were acquired using the scanning probe 
microscopy unit of the same setup, with a resolution of
 1024~$\times$~1024 pixels per scan 
 and NSG10 probes (NT-MDT). 
The AFM and optical microspectroscopy fields of view were 
aligned using laser confocal imaging of the AFM probe tip. 
All Raman and SERS spectra were collected after single-PMS 
AFM mapping. This confocal alignment procedure ensured 
that each spectrum of an individual PMS could be directly 
correlated with its AFM image. 
The bulk PMS Raman spectrum was obtained by drying the PMS stock
solution on a glass substrate, followed by Raman measurements
performed in the visible ``coffee-ring'' region. 
Single-PMS non-enhanced Raman spectra were acquired in 
the same manner as single-PMS SERS spectra, using the combined
AFM/microspectroscopy system. In this case, individual PMSs 
were deposited on a bare glass substrate, and subtraction of 
the glass Raman spectrum was performed. 
 \subsection{Mie Expansion of the Electromagnetic Field in a Core--Shell Microsphere}
 \label{Sec Mie expansion of EM field in core--shell microsphere}
 The electromagnetic field in a core--shell resonator, namely,
 a metal sphere filled with Raman-active
 {molecules,} can be found using the 
 Mie expansion \cite{Mie1908,AdenKerker1951,
SarychevIvanovBykov2023DD,SarychevIvanovBykov2023QE,
SarychevIvanovBykov2024QE,SarychevIvanovBykov2024ModEl}. 
 Let us briefly recapitulate the main points of this method. 
 We assume that monochromatic light with frequency
 $\omega$ and amplitude 
 equal to unity is incident on {a} core--shell 
 sphere along the \textit{z}-axis. 
 Below the time dependence $\exp(-i\omega t)$ is used, and
 media suppose to be nonmagnetic.
 The electric field of the light
 is polarized along {the} \textit{x}-axis. 
 In this section we measure all the distances in terms of the inverse 
 wavevector $ 1/k = \lambda /2 \pi = c/\omega $ and 
 measure all the fields in terms of the amplitude
 $ E_{0} $ of the incident plane EM wave.
 Then the electric and magnetic fields of the incident light {are}
 \begin{eqnarray} 
 \label{E0Wave}
 \mathbf{E}_{0} = \hat{\mathbf{x} } \exp (i n_{3} z ) = 
 \hat{\mathbf{x} }\,	\sum_{ {q = 0}}^{\infty}
 i^{q} (2 q + 1) P_{q} (\cos \theta) j_{q}( n_{3} r),
 \\
 \label{H0Wave}
 \mathbf{ H}_{0 } =\hat{\mathbf{y} } \, n_{3} 
 \exp (i n_{3} z ) = \hat{\mathbf{y} } \, n_{3} \, \sum_{ {q = 0}}^{\infty} 
 i^{q} (2 q + 1) P_{q} (\cos \theta) j_{q} ( n_{3} r ),
 \end{eqnarray}
where $ \hat{\mathbf{x}}$ and $ \hat{\mathbf{y}} $ are unit vectors,
$ j_{q}( n_{3} r) $ and $ P_{q}(\cos \theta ) $ 
are the spherical Bessel function 
and the Legendre polynomial of order $q${, respectively}, 
$ \theta $ is the polar angle measured from the \textit{z}-axis, 
$ n_{3} = \sqrt{\varepsilon_{e}} $ 
is the refractive index of the ambient medium. 
Eqs.\,(\ref{E0Wave}) and (\ref{H0Wave}) can be verified by direct 
computer simulation or by comparing {the} coefficients of the term 
$ \left( n_{3} z \right)^{q} \
\equiv
 \left( n_{3} r \cos \theta \right)^{q} $ 
{on the left- and right-hand sides} 
of Eqs.\,(\ref{E0Wave}) and (\ref{H0Wave}) 
(see e.g. \cite{LandauLifshitzQuantum1991}, Sec.\,34).
A solution of {Maxwell's equations} is typically simplified
 {through the use} of a vector potential.
The standard method {uses a} vector potential that
 {has} the same symmetry as the problem under consideration.
The electromagnetic field scattered by a sphere is found using
electric vector potentials of
 \begin{equation} 
 \label{EleSpPot}
 \left\lbrace \mathbf{M}_{q, j}^{(i)} (\mathbf{r}), \, 
 \mathbf{M}_{q,h}^{(i)} (\mathbf{r}) \right\rbrace 
 = 
 \mathbf{r} \, w_{q} n_{i} \cos (\varphi ) P_{q}^{1} (\cos \theta ) 
 \left\lbrace j_{q} (n_{i} r), h_{q} (n_{i} r) \right\rbrace 
 \end{equation}
and magnetic vector potentials
 \begin{equation}
 \label{MagSpPot}
 \left\lbrace \mathbf{N}_{q, j}^{(i)} (\mathbf{r}), \,
 \mathbf{N}_{q, h}^{ (i)} (\mathbf{r}) \right\rbrace 
 = -\mathbf{r } \, w_{q} \sin ( \varphi) P_{q}^{1} (\cos \theta ) 
 \left\lbrace j_{q} (n_{i} r), h_{q} (n_{i} r) \right\rbrace, 
 \end{equation}
composed of spherical harmonics
from Eqs.\,(\ref{E0Wave}) and (\ref{H0Wave}),
where $ q = 1,2,3 \ldots $, $ \varphi$ and $ \theta $ 
are the azimuthal and polar angles{, respectively}, 
the index $i = 1, 2, 3$ 
denotes the internal dielectric, 
{metal shell, and the} ambient medium; 
$n_{1} = \sqrt{\varepsilon_d} $, 
$n_{2} = \sqrt{\varepsilon_{ m } }$, 
and $ n_{3} = \sqrt{\varepsilon_{ e } }$
are the refractive indices of the internal dielectric, 
 {metal shell, and the} surrounding space, respectively, 
$P^{1}_{q} (\cos \theta ) = \partial P_{q} (\cos \theta ) / \partial \theta$
is the associated Legendre polynomial of first degree and order $ q$; 
functions $j_{q}$ and $h_{q}$ are the spherical Bessel 
and Hankel functions of the first kind, 
respectively{; and the coefficient is}
$ {w_{q} = i^{q}(2q + 1) /q(q + 1)}$. 

It is easy to check {that the} four vector functions 
$ \mathbf{curl} \,\mathbf{ M}_{q, j}^{(i)} (\mathbf{r}) $, 
$ \mathbf{curl} \,\mathbf{ M}_{q, h}^{(i)} (\mathbf{r})$, 
$ \mathbf{curl} \,\mathbf{ N }_{q, j}^{(i)} (\mathbf{r}) $
and 
$ \mathbf{curl} \,\mathbf{N}_{q, h}^{(i)} (\mathbf{r})$
are solutions of the wave equation, 
which can be verified by direct calculation.
Therefore, functions $ \mathbf{M} $ and $ \mathbf{N} $ can be used 
as vector potentials to construct 
a solution of the Maxwell equations. 
 {There} are two solutions due to the symmetry of {Maxwell's}
equations with respect to the mutual exchange between 
\textbf{E} and \textbf{H}.
The ``electric'' and ``magnetic'' solutions {are}
 \begin{align}
 & \left\lbrace 
 \mathbf{H}_{\text{E}, q, j}^{(i)}, 
 \mathbf{H}_{\text{E}, q, h}^{(i)}
 \right\rbrace
 = 
 \mathbf{curl} 
 \left\lbrace 
 \mathbf{ M}_{q, j}^{(i)}, \mathbf{ M}_{q, h}^{(i)} 
 \right\rbrace, \nonumber \\ \label{EHfromM}
 & \left\lbrace 
 \mathbf{E}_{\text{E}, q, j}^{(i)}, \mathbf{E}_{\text{E}, q, h}^{(i)}
 \right\rbrace = 
 \frac{i}{ n_{i}^{2}} \mathbf{curl} 
 \left\lbrace \mathbf{ H}_{\text{E}, q, j}^{(i)}, \mathbf{ H}_{\text{E}, q, h}^{(i)}
 \right\rbrace;
 \end{align}
 and 
 \begin{align} 
 & \left\lbrace 
 \mathbf{E}_{\text{H}, q, j}^{(i)}, \mathbf{E}_{\text{H}, q, h}^{(i)}
 \right\rbrace = \mathbf{curl} 
 \left\lbrace \mathbf{N}_{q, j}^{(i)}, {\mathbf{N}_{q, h}^{(i)}} 
 \right\rbrace, \nonumber \\ \label{EHfromN} 
 & \left\lbrace 
 \mathbf{H}_{\text{H}, q, j}^{(i)}, \mathbf{H}_{\text{H}, q, h}^{(i)}
 \right\rbrace = 
 - i \, \mathbf{curl} 
 \left\lbrace \mathbf{E}_{\text{H}, q, j}^{(i)}, \mathbf{E}_{\text{H}, q, h}^{(i)}
 \right\rbrace,
 \end{align}
 respectively.
 Four EM fields
 $ \mathbf{E}_{\text{E}, q, j}^{(i)}, \mathbf{H}_{\text{E}, q, j}^{(i)}, 
 \mathbf{E}_{\text{H}, q, j}^{(i)} $, 
 and 
 $ \mathbf{H}_{\text{H}, q, j}^{(i)}, $
 include the spherical Bessel functions and, therefore, these fields 
 are regular at the origin $ r = 0 $. 
 The incident light 
 $ \left\lbrace \mathbf{E}_{0}, \mathbf{H}_{0} \right\rbrace $
 can be {expressed as a sum} of these fields. Projecting
 the incident plane wave, given by Eqs.\,(\ref{E0Wave}) and 
 (\ref{H0Wave}), {onto} the fields (\ref{EHfromM}) and
 (\ref{EHfromN}){,} we obtain
 \begin{align} 
 	 \label{E0WaveMN}
 &\mathbf{E}_{0} = \mathbf{E}_{0\text{E}} + \mathbf{E}_{0\text{H}} =
 \sum_{q = 1}^{\infty} \mathbf{E}_{\text{E}, q, j}^{(3)}+ 
 \sum_{q = 1}^{\infty} \mathbf{E}_{\text{H}, q, j}^{(3)},
 \\
 \label{H0WaveMN} 
 &\mathbf{ H}_{0 } = \mathbf{H}_{0\text{E}} + \mathbf{H}_{0\text{H}} =
 \sum_{q = 1}^{\infty} \mathbf{H}_{\text{E}, q, j}^{(3)}+ 
 \sum_{q = 1}^{\infty} \mathbf{H}_{\text{H}, q, j}^{(3)}
 \end{align}
The electromagnetic fields given by Eqs.\,(\ref{EHfromM}) 
and (\ref{EHfromN}) can be used to construct a solution 
 {to} Maxwell's equations in the dielectric core, metal shell, and
 {the surrounding space}.
 {Thus, the following equations are obtained for} 
 the {internal} electric field 
 $ \mathbf{E}_{in} $, shell field $ \mathbf{E}_{m} $, 
 and {the total external field} $ \mathbf{E}_{ex} $
 \begin{align}
 \label{Ein}
 & \mathbf{E}_{in}=\sum _{q=1}^{\infty } 
 \left(
 A_{\text{E} q} \mathbf{E}_{\text{E},q,j}^{(1)}+
 A_{\text{H} q} \mathbf{E}_{\text{H},q,j}^{(1)}
 \right),
 \\ 	\label{Em}
 & \mathbf{E}_{m}=\sum _{q=1}^{\infty } \left( 
 B_{\text{E} q} \mathbf{E}^{(2)}_{\text{E}, q, j} + 
 B_{\text{H} q} \mathbf{E}^{(2)}_{\text{H}, q, j} + 
 C_{\text{E} q} \mathbf{E}^{(2)}_{\text{E}, q, h}+
 C_{\text{H} q} \mathbf{E}^{(2)}_{\text{H}, q, h} 
 \right),
 \\ 	\label{Eex}
 & \mathbf{E}_{ex}= {\mathbf{E}_{0}} +
 \sum _{q=1}^{\infty } \left(
 D_{\text{E} q} \mathbf{E}^{(3)}_{\text{E}, q, h}+
 D_{\text{H} q} \mathbf{E}^{(3)}_{\text{H}, q, h}
 \right),
 \end{align}
where the "electric" coefficients 
$A_{\text{E} q}, B_{\text{E} q} $, $C_{\text{E} q}, D_{\text{E} q}$
and "magnetic" coefficients 
$ A_{\text{H} q}, B_{\text{H} q}$, $C_{\text{H} q}, D_{\text{H} q} $
 define the EM field in the whole system.
The linear equations for the coefficients 
are found by matching electric and magnetic fields
at the interface between the dielectric core and the metal shell 
and at the interface between the metal shell and {the} ambient medium.
The electric 
$ \mathbf{E}_{\text{E}, q, j}^{(i)}, \;
\mathbf{E}_{\text{E}, q, h}^{(i)}$ and magnetic
$ \mathbf{H}_{\text{E}, q, j}^{(i)}, \;
\mathbf{E}_{\text{E}, q, h}^{(i)}$
fields
are constructed in such a way that they have the same angular
 dependence {in the}
core ($ i=1 $), metal shell ($ i=2 $), and surrounding space ($ i=3 $).
Therefore{,} only the radial parts should be matched.
 {The same is true} for the other fields.
Equations for the "electric" 
coefficients have the following simple {form, which includes the radial} 
Bessel $j_{q}$ and Hankel $h_{q}$ functions:
\begin{align} 
& n_1 A_{\text{E} q} j_{a1}
 =
n_2 (B_{\text{E} q} j_{a2} + C_{\text{E} q} h_{a2} ),
\nonumber
\\
& n_2 A_{Eq} j'_{a1}
= n_1 \left(B_{Eq} j'_{a2} + C_{Eq} h'_{a2} \right),
\label{EqsABCD}
\\
&n_2 \left(B_{Eq} j_{d2}+C_{Eq} h_{d2}\right) =
n_3 \left(D_{Eq} h_{d3}+j_{d3}\right), 
\nonumber
\\
&n_3 \left(B_{Eq} j'_{d2}+C_{Eq} h'_{d2}\right)
=
n_2 \left(D_{Eq} h'_{d3}+j'_{d3}\right),
\nonumber
\end{align}
where 
$ j_{ai} = j_q \left( n_{i} a \right)$,
$ j'_{ai} = (q+1) j_q \left( n_{i} a \right) - 
\left( n_{i} a \right) j_{q+1} \left( n_{i} a \right),$
and
$ h_{ai} = h_q \left( n_{i} a \right)$, 
$ h'_{ai} = (q+1) h_q \left( n_{i} a \right) 
- \left( n_{i} a \right) h_{q+1}\left( n_{i} a \right), $
 {where we recall that} all the distances are 
measured in terms of $1/k = c / \omega $.
The determinant of Eqs.\,(\ref{EqsABCD}) {is}
\begin{align} \label{DetE}
{\rm Det_{E}} =
\left(n_1^2 j_{a1} j'_{a2}-n_2^2 j'_{a1} j_{a2}\right) 
\left(n_2^2 h_{d2} h'_{d3}-n_3^2 h'_{d2} h_{d3}\right)+ 
\nonumber \\
\left(n_2^2 j'_{a1} h_{a2}-n_1^2 j_{a1} h'_{a2}\right)
\left(n_2^2 j_{d2} h'_{d3}-n_3^2 j'_{d2} h_{d3}\right)
\end{align} 
The determinant does not depend on the amplitude of the incident wave.
Therefore, it gives the intrinsic plasmonic properties 
of the considered core--shell
structure. {The zeros} of the determinant 
(\ref{DetE}) define the ``electric'' resonance frequencies 
$ \omega_{E, q} (a, d)$ 
in  the  core--shell microsphere. 
$ \Re\omega_{E, 1} (a, d)$ is
the frequency of the dipole resonance, $ \Re\omega_{E, 2} (a, d)$
is the {quadrupole} resonance{,} and so on. 

 {Similar} equations can be obtained for the ``magnetic'' coefficients 
$ A_{\text{H} q}, B_{\text{H} q}, C_{\text{H} q} $, 
and $ D_{\text{H} q} $ in Eqs.\,(\ref{Ein}), (\ref{Em}), 
and (\ref{Eex}).
The corresponding ``magnetic'' determinant {is}
\begin{align} \label{DetH}
{\rm Det_{H} } =
\left(j'_{a1} j_{a2}-j_{a1} j'_{a2}\right) 
\left(h'_{d2} h_{d3}-h_{d2} {h'_{d3}}\right) - 
\nonumber \\
\left(j_{a1} h'_{a2}-j'_{a1} h_{a2}\right) 
\left(j_{d2} h'_{d3}-j'_{d2} h_{d3}\right)
\end{align} 
 {The zeros} of the determinant (\ref{DetH})
  {define the ``magnetic'' resonance frequencies} 
$ \omega_{H, q} (a, d )$ 
in   {a} core--shell metal microsphere. These resonances are well
developed when the   {losses} in   {the} metal   {are} small. The coefficients 
$ A_{\text{E} q} $ 
and 
$ A_{\text{H} q}$ that define the internal electric field in Eq.\,(\ref{Ein})
  {are}
\begin{equation}
\label{AeAh}
A_{\text{E} q} = \frac{n_1 n_2}{ a (a+d) \, {\rm Det_{E}} }, 
\;\;
A_{\text{H} q} = \frac{1}{a n_2 n_3 (a+d) \, {\rm Det_{H}} }.
\end{equation}
The coefficients $ D_{\text{E} q} $ and 
$ D_{\text{H} q}$ that define the external electric field in
Eq.\,(\ref{Eex})
  {are}
\begin{align}
\label{De}
D_{\text{E} q} = &\frac{1}{ {\rm Det_{E}}}
\left[ n_1^2 n_2^2 j_{a1} j'_{d3} \left(h'_{a2} j_{d2} - 
 j'_{a2} h_{d2}\right) + \right. 
\nonumber \\
&n_2^4 j'_{a1} j'_{d3} \left(j_{a2} h_{d2} - 
h_{a2} j_{d2}\right) + 
n_1^2 n_3^2 j_{a1} j_{d3} \left(j'_{a2} h'_{d2} 
- h'_{a2} j'_{d2}\right)+
\nonumber\\ 
& \left. n_2^2 n_3^2 j'_{a1} j_{d3} 
\left( h_{a2} j'_{d2} -j_{a2} h'_{d2}\right)
\right] 
\end{align}
\begin{align}
D_{\text{H} q} = & \frac{1}{ {\rm Det_{H}}}
\left[ 
\left(j'_{a1} j_{a2}-j_{a1} j'_{a2}\right) \left(h_{d2}
j'_{d3}-h'_{d2} j_{d3}\right)-
\right. \nonumber \\ 
& \left. 
\left(j'_{d2} j_{d3}-j_{d2}
	j'_{d3}\right) \left(j_{a1} h'_{a2}-j'_{a1} h_{a2}\right)\right] 
 \label{Dh}
 \end{align}
Equations (\ref{Ein}), (\ref{Eex}), (\ref{DetE}), (\ref{DetH}), 
(\ref{AeAh}), (\ref{De}) and (\ref{Dh})
fully define the EM field inside and outside 
 {the} core--shell sphere{,}
including {the} local {electric-field} 
enhancement that {gives rise to the} 
SERS phenomenon. 
In a real SERS experiment the laser frequency $\omega$
is typically fixed by the laser and the {entire optical} system. 
Very often{, lasers with wavelengths of} 
$532\,nm$ and $785\,nm$ 
are used in Raman devices.
A plasmon resonance in a core-shell system can be achieved simply 
by tuning the thickness $d$ of the metal shell.
The thickness can be easily 
changed in the process of the core--shell preparation (see Sec.\,
 \ref{SubSec Synthesis of PMS and fabrication of silver substrates}). 
\begin{figure}[!htbp]
	\centering
\includegraphics[ width=\textwidth]{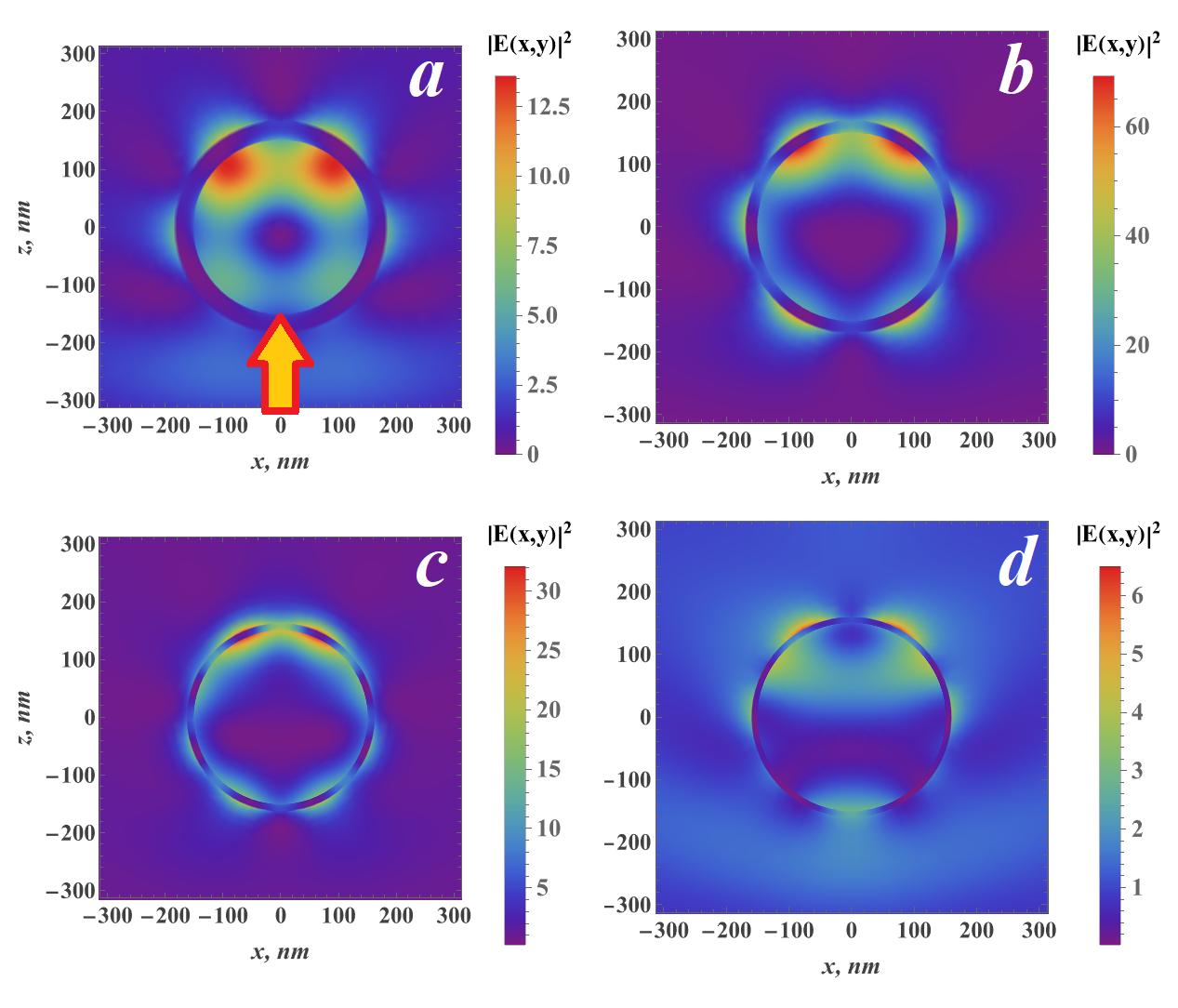}
	\caption{
		Distribution of {the} electric field 
 $\left| E (x,z)\right|^{2} $ {in the} plane $ y = 0 $
in {a} dielectric sphere 
($ n_d=1.6; \; \varepsilon_{d}=n_d^2=2.56 $) with radius
$ a = 150 \, nm$, which is surrounded by {a} silver shell
with {thicknesses} $ a) \, d = 30.5 $, $ b) \, d = 17.8 $, 
$ c) \, d = 11.5 $, $ d) \, d = 8.8\;nm$; 
 {an $x$-polarized EM plane wave} with $\lambda = 550\;nm$ is
 incident along {the $z$-axis, i.e., from below; the
direction of the incident light is shown by the red-yellow arrow.}}
	\label{Fig1}
\end{figure}
When the materials and radius of the core are fixed
 the electric plasmon resonances
correspond to the minima of the determinant 
${\rm Det_{E}}(d) $ as {a} function of the shell thickness $d$.
The {resonant} metal thickness $ d_{res,q} (a, \lambda)$ 
is found from the dispersion equation 
\begin{equation}
	\label{DispE}
	 {d_{res,q}(a,\lambda)=
	\underset{ d }{\operatorname{arg\,min}}\,
	\left|{\rm Det_{E}}(q,\lambda,a,d)\right|},
\end{equation}
where $ q$ is the order of the 
electric plasmon resonances in the dielectric 
sphere covered by the metal shell.
When the core radius $ a $ is smaller than the wavelength $\lambda$
the cylinder functions can be {expanded} 
in {a} series {in} $a/\lambda$.
Then {the} solution of the dispersion Eq.\,(\ref{DispE}) 
gives the resonance thickness of the metal shell
\begin{equation}
 \label{dRes}
 \frac{d_{res,q}}{a} \simeq
 \Re
 \left\lbrace \left[ \frac{q (q+1)
 \left(\varepsilon_m-\varepsilon_d\right)
 \left(\varepsilon_m-\varepsilon_e\right)}
 {\left(q \left(\varepsilon_d+\varepsilon
 _m\right)+\varepsilon_m\right) 
 \left(\varepsilon_e+q \left(\varepsilon_e+
 \varepsilon_m\right)\right)}\right]^{\frac{1}{2 q+1}}
 \right\rbrace -1,
\end{equation}
where we suppose that metal permittivity 
$ \varepsilon_{m} = \varepsilon'_{m}+ i \varepsilon''_{m}$
is mainly negative, i.e.,
$ \left| \varepsilon'_{m} \right| \gg \varepsilon''_{m}$,
it is also supposed that 
$\left| \varepsilon_{m} \right| > \varepsilon_{d},
 (1+q^{-1}) \varepsilon_{e}$.
Expression in square brackets in Eq.\,(\ref{dRes}) is
not one-valued function.
 Their value is supposed chosen in accordance 
 with the condition that the complex quantity which is raised to 
the power $ (2 q +1) ^{-1} $ is taken with the argument 
whose absolute value is least.
 
\begin{figure}[!htbp]
 	\centering
 	\includegraphics[ width=\textwidth]{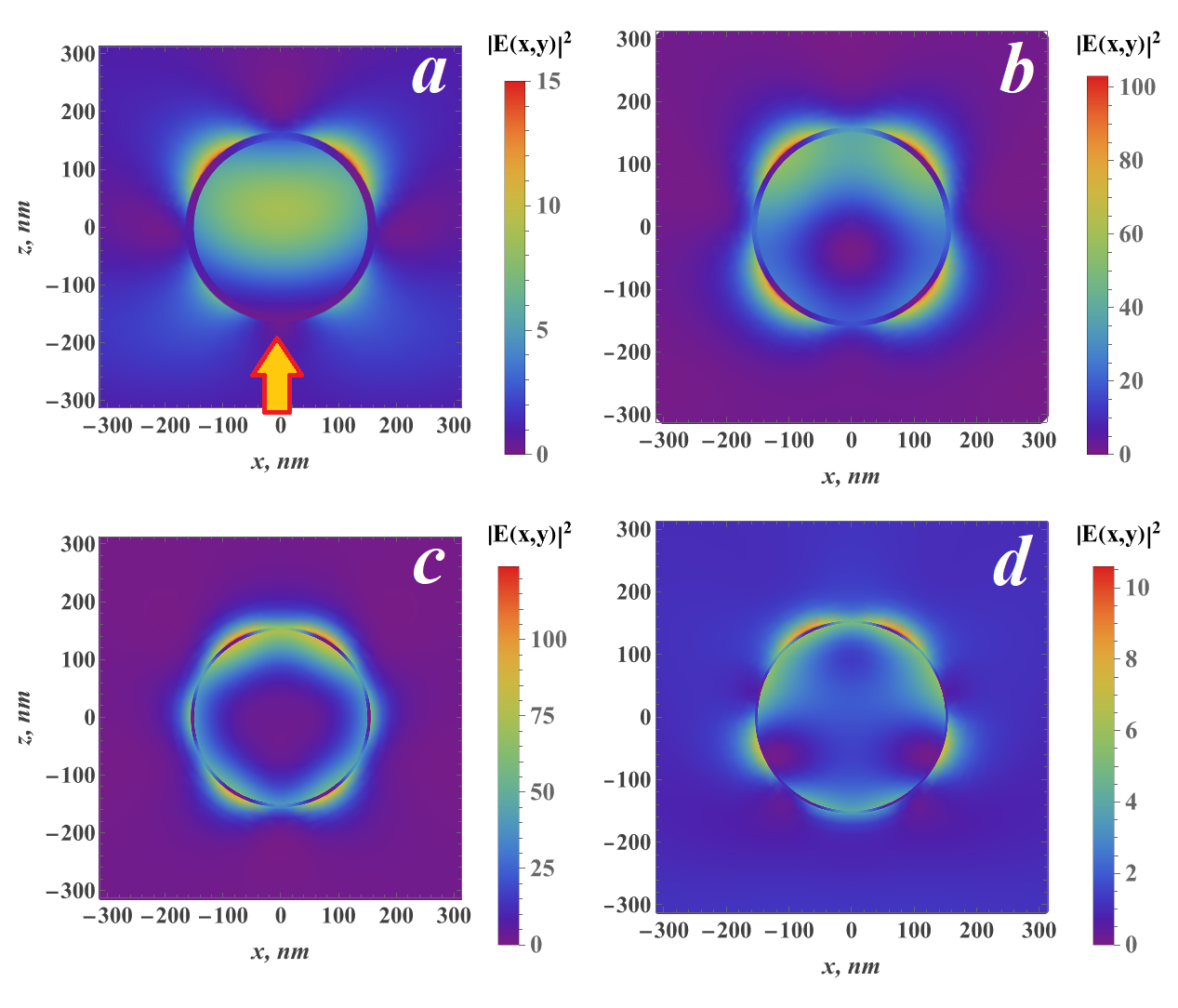}
	\caption{ 
		Distribution of {the} local electric field 
 $\left| E (x,z)\right|^{2} $ {in the} plane $ y = 0 $
 in {a} dielectric sphere 
 ($ {n_d=1.6; \; \varepsilon_{d}=n_d^2=2.56}$) 
 with radius
 $ a = 150 \, nm$, which is surrounded by {a} silver shell
 with {thicknesses} $ a) \, d = 13.8 $, $ b) \, d = 8.7 $, 
 $ c) \, d = 5.4 $, and $ d) \, d = 4.0 \; nm$; 
	 {an $x$-polarized EM plane wave} with $\lambda = 800\, nm$ is
	incident along {the $z$-axis}, as in Fig.\, \ref{Fig1}; the
	direction of the incident light is shown by the red-yellow arrow.}
 	\label{Fig2}
 \end{figure}

The quasistatic equation (\ref{dRes}) holds when 
$a n_{1} k/q \ll 1$, $a n_{3} k/q \ll 1$,
and the shell thickness $d $ is less than the skin depth 
$\delta_{sk}\gg d$ (see discussion in 
Sec.\, \ref{Sec Results and Discussion}).
In the red and infrared optical bands the metal permittivity
 $\varepsilon_{m}$ is negative but large in absolute value 
 $ \left| \varepsilon_{m} \right| \gg 1$. Then Eq.\,(\ref{dRes})
simplifies to 
\begin{equation}
	\label{dRes0}
	d_{\text{res,q}}\sim
	-\frac{a \left(\dfrac{q \varepsilon_d}{q+1}+\varepsilon_e\right)}
	{q \,	\varepsilon'_m}
	 {\xrightarrow{\ q\gg1\ }}
	 -\frac{a \left(\varepsilon_d + \varepsilon_e \right)}{q \, \varepsilon'_m},
\end{equation}
where the real part $\varepsilon'_m $ of the metal permittivity is 
assumed to be negative.
Simple estimates (\ref{dRes}) and (\ref{dRes0}) are in qualitative
 agreement with {the structure} of the {plasmon} fields shown 
 in Figs.\, \ref{Fig1} and  \ref{Fig2}.
 
We consider {a} PMS with {a} core radius $ a = 150\,nm $
and various silver shell {thicknesses} $ d $ 
that {yield resonances} at a fixed laser frequency.
For the core radius $ a = 150\,nm $ and laser wavelength
$ \lambda = 550 \,nm $ 
the {thicknesses} $ d_{res,q}$ of the {resonant} silver shell 
 {are} 
$ d_{res,q} = 30.5, \, 17.8, \,11.5, \, 8.8, \, 7.2, \, 6.1,\ldots \, nm$;
for $ q=2,3,4,\ldots ${.}
For the same core radius $ a = 150\,nm$ and 
$ \lambda = 800 \,nm ${,} the {resonant silver-shell thicknesses 
are} 
$ d_{res,q} = 
13.8, \, 8.7, \, 5.4, \, 4.0, \, 3.2, \, 2.7, \, 2.3,\ldots \, nm$
for $ q = 1, 2,3, \ldots $.
Dipole, {quadrupole}, hexapole, and octapole resonances 
are shown in Figs.\, \ref{Fig1} and  \ref{Fig2}.
 {Note that} the dipole resonance is {damped} for 
 $ \lambda =550 \, nm$ due to the radiation loss.
 
\begin{figure}[!htbp]
 	 \centering
 \hspace{-0.8 cm} 
 	\includegraphics[width=\textwidth]{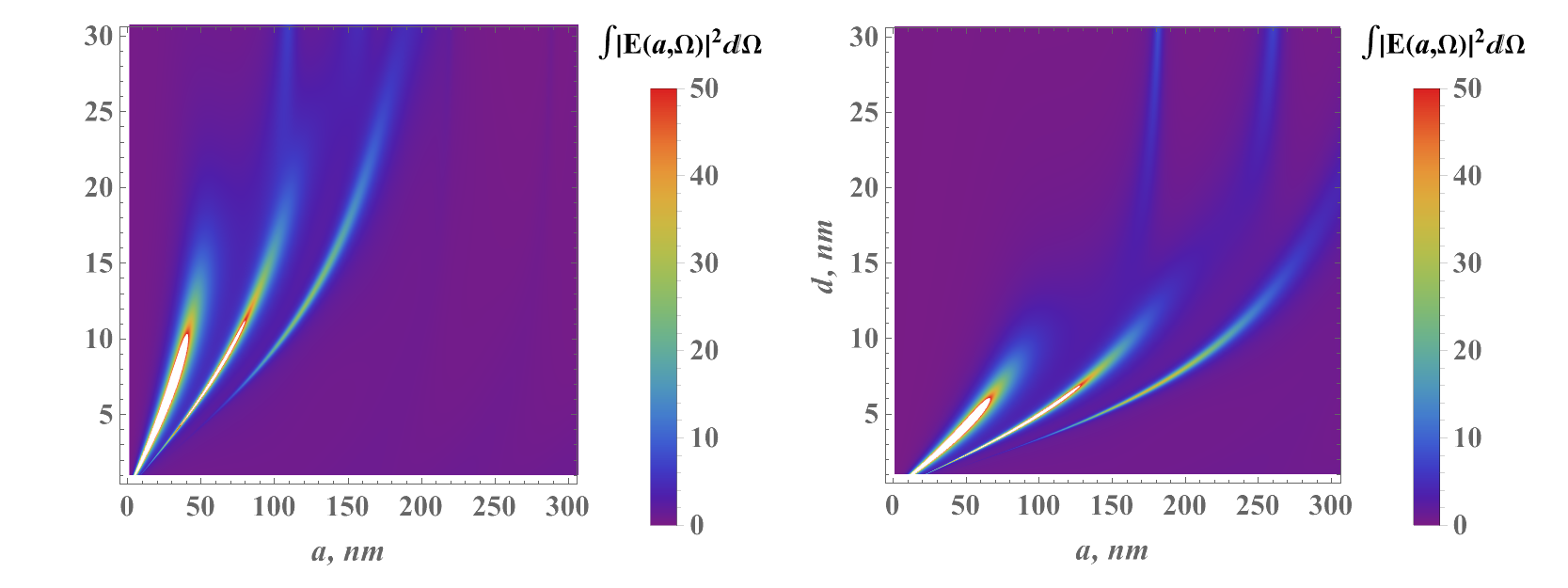}
 \caption{ 
 	 {The averaged intensity} 	
$\left\langle \left| E (x,z) \right|^{2} \right\rangle $ 
 of the local electric field in a silver-dielectric microsphere 
 with a dielectric core and a silver shell as a function 
 of the radius $a$ of the dielectric core 
 ($ n_d=1.6; \; \varepsilon_d=n_d^2=2.56$) 
 and the thickness $d$ of the silver shell{. The} 
 intensity of the local electric field is averaged over 
 the inner surface of the metal shell{. The wavelength of} 
 the incident light {is} $\lambda = 550 \, nm$ 
 {in panel (a)} and $ \lambda = 800 \, nm$ {in panel (b)}.}
 	\label{Fig3}
 \end{figure}

 {To gain insight into} the electric field enhancement at the plasmon 
resonance{,} we obtain explicit analytical equations for the intensity 
of the local electric field 
$\left\langle \left| E \right|^{2} \right\rangle_{in} $ and
$\left\langle \left| E \right|^{2} \right\rangle_{ex} $
which are averaged over 
the internal and external {surfaces} of the metal shell{,}
 {respectively,} 
 \begin{align}
&\left\langle \left| E \right|^{2} \right\rangle_{in} = 
 \frac{1}{4 \pi }\oint \left| E_{in}\left( a, \theta , \varphi \right)
 \right|^{2} d\,\Omega =
 \nonumber \\
& \sum_{q =1}^{\infty} \frac{(2 q+1)}{2} \left( 
 \frac{q (q+1) {\left|j_{a1}\right|^2}+ {\left|j'_{a1}\right|^{2}} }
 { {\varepsilon_{d} a^2} } |A_{\text{E} q }| ^2
 + {\left|j_{a1}\right|^2} \left| A_{\text{H}q}\right|^2 \right) ,
\end{align}
 \begin{align}
 &\left\langle \left| E \right|^{2} \right\rangle_{ex} =
 \frac{1}{4 \pi }\oint \left| E_{ex}
 \left( a+d, \theta ,\varphi \right) \right|^{2}
 d\,\Omega =
 \nonumber \\
&\sum_{q =1}^{\infty}
 \frac{2 q+1}{2} \left(
 \frac{ q (q+1) D_{1}+D_{2}}{ { \varepsilon_{e }(a+d)^2}}+
 \left| D_{\text{H} q} h_{d3}+j_{d3}\right|^{2} \right),
 \end{align}
 where 
 $d\Omega = \sin \theta \,d \theta \, d \varphi $ is the element of 
 the solid angle, the coefficients 
 $ j_{a1}$, $ j_{d3}$, $ h_{d3 }$,
 $ j'_{a1}$, $ j'_{d3}$, and $ h'_{d3 }$ 
 are define at Eqs.\,(\ref{EqsABCD}); 
 $D_1 = \left| D_{\text{E}q } h_{d3}+j_{d3} \right| ^{2}$, and
 $D_2=\left| D_{\text{E}q} h'_{d3}+j'_{d3}\right|^2$;
 the coefficients $A_{\text{E} q }$, $A_{\text{H} q }${,}
 $D_{\text{E} q }${,} and $D_{\text{H} q }$
 are given by Eqs.\,(\ref{AeAh}), (\ref{De}) and (\ref{Dh}). 
 The averaged {internal intensity}
 $\left\langle \left| E \right|^{2} \right\rangle_{in}$
 is shown in Fig.\,\ref{Fig3} as {a} function of the radius $a$ 
 of the dielectric core 
 and the thickness $d$ of the silver shell. 
\subsection{Numerical simulation of {a} PMS on 
 {a} silver substrate}
 \label{SubSec Numerical simulation of PMS on 
 	silver substrate}
The amplitude of the local electric field in the polystyrene microsphere 
was obtained by directly solving Maxwell's equations in COMSOL 
Multiphysics. A plane electromagnetic wave with a wavelength of 
$\lambda = 532$~nm was incident normally from the top
(wave vector $\mathbf{k}$ 
directed along $-z$) on a dielectric sphere of diameter 
$ 2 a = 300 \, nm$
resting on a smooth metal substrate of thickness $t = 200$~nm. The refractive indices used in the model were $n = 1$ for air and 
$n = 1.6 $ for polystyrene~\cite{SultanovaKasarovaNikolov2009}. 
The optical constants of silver were taken from experimental 
data~\cite{JohnsonChristy1972}. 
\begin{figure}[!htbp]
	\centering
	\includegraphics[width=0.7\textwidth]{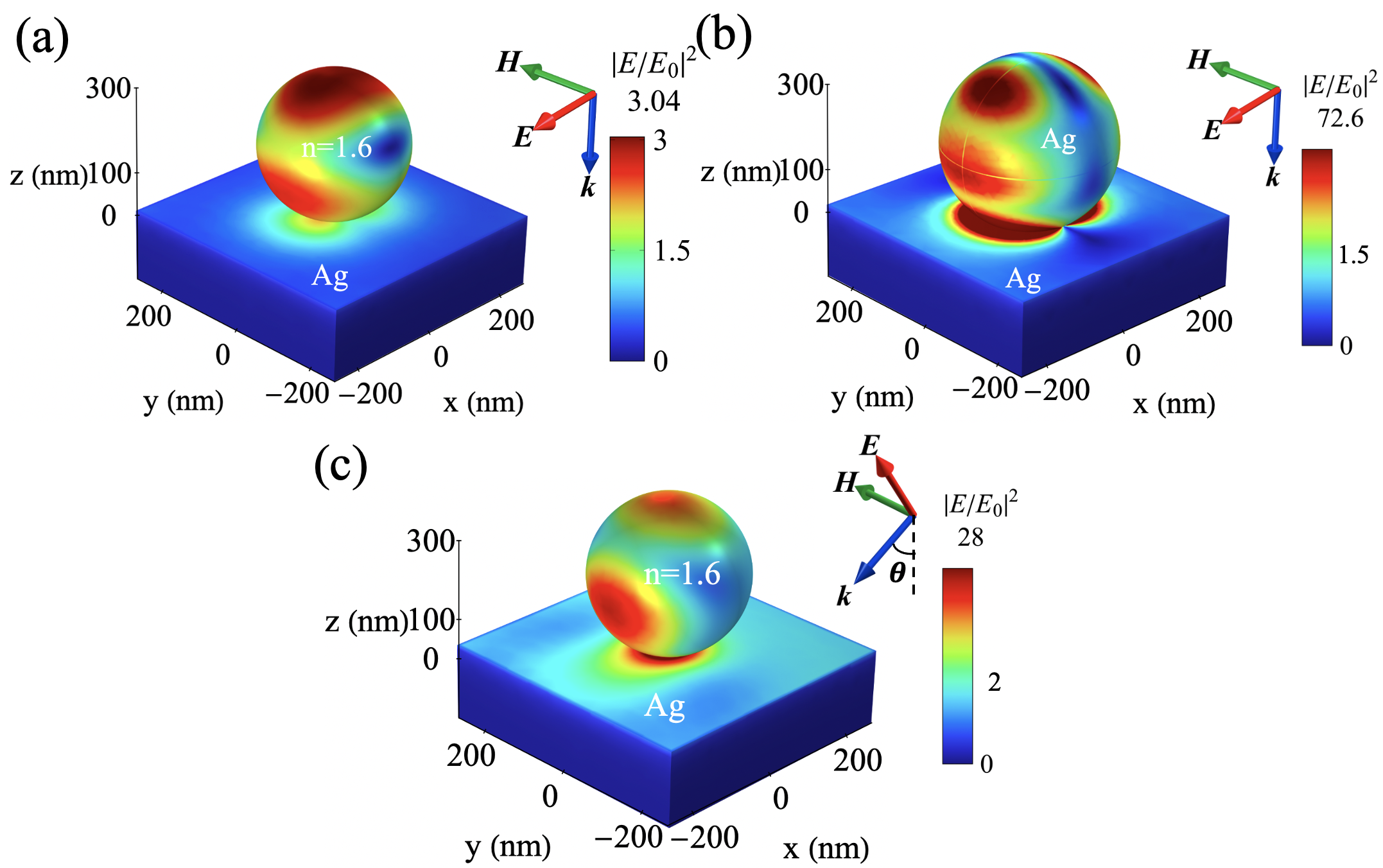}
	\caption{
		Three-dimensional distributions of the normalized electric-field 
		intensity, $|E/E_0|^2$, around a polystyrene microsphere (PMS) 
		supported by a planar silver substrate at $\lambda=532\,\mathrm{nm}$. 
		(a) Bare PMS with refractive index $n=1.6$ under normal incidence. 
		(b) PMS coated with a $d=20\,\mathrm{nm}$ silver shell under the 
		same illumination conditions. In panels (a) and (b), the incident 
		electric field is polarized along the $x$ axis. 
		(c) Bare PMS illuminated by a TM-polarized plane wave incident at 
		$\theta=50^\circ$ relative to the substrate normal; the wave vector 
		$\mathbf{k}$ and electric-field vector $\mathbf{E}$ lie in the 
		$xz$ plane. The sphere diameter is $2a=300\,\mathrm{nm}$ in all cases. 
	}
	\label{Fig4}
\end{figure}
Following the approach described in 
 {Ref.~\cite{COMSOL},} 
the simulation employed two study steps. First, the background 
field of the plane wave incident on the bare substrate was computed. 
Then, the total electromagnetic field was determined 
with the dielectric sphere present{, as shown} in
 Figs.\, \ref{Fig4}(a-c).
\section{Results and Discussion}
\label{Sec Results and Discussion}
\subsection{Plasmon resonances in core--shell microparticle and dark modes}
\label{SubSec Plasmon resonances in core--shell microparticle
and dark modes}
It is generally accepted that the local electric field is 
significantly enhanced due to {a} plasmon resonance
in a metal nanoparticle a few nanometers in diameter
or in an interparticle gap of the same size.
Less obvious is that the internal field can be significantly
larger than the incident field when the diameter
of a core--shell particle is equal to or even greater
than the wavelength of the incident light.
This effect is most pronounced when the frequency of light
$\omega$ 
is close to the plasmon resonance. 
 {In a dielectric microsphere with radius $a$ 
surrounded by a metal shell with thickness $ d \ll a$, the core--shell plasmon is excited by the incident light.} 
The permittivity of the metal
$ \varepsilon_{ m } = \varepsilon '_{ m } + \varepsilon ''_{ m }$ 
is mostly negative for good optical metals such as silver or gold, 
where 
$\varepsilon '_{ m } < 0 $ 
and 
$ \varepsilon ''_{ m } \ll \left| \varepsilon_{ m } \right|$. 
The sphere is filled with {a dielectric of permittivity} 
$\varepsilon_{d}$. 
The thickness $d$ of the metal shell is chosen so that
 {the} incident light somehow penetrates the metal film
surrounding the dielectric core. {Note that} light can also penetrate
into the core through holes in the metal shell \cite{SilvaNygaDrachev2015}.

A propagating surface plasmon (SP)—an electromagnetic 
field coupled to electric charges—is an intrinsic property 
of a metallic nanofilm.
 {A symmetric SP} propagates in a flat metal 
nanofilm. The surface charges of this SP have the same {sign} 
on both sides of the film{; therefore, the} 
electric field {is concentrated mainly}
outside the film. The wave vector 
$k_{p}$ of the symmetric SP {is} 
\begin{equation} \label{kp} 
k_{p}
\simeq \frac{1}{2d}\ln\!\left[
\frac{(\varepsilon_m-\varepsilon_d)(\varepsilon_m-\varepsilon_e)}
{(\varepsilon_m+\varepsilon_d)(\varepsilon_m+\varepsilon_e)}\right]
 \simeq -\frac{\varepsilon_d+\varepsilon_e}{d\,\varepsilon_m},
\end{equation}
where $\varepsilon_{d}$ and $\varepsilon_{e}$ 
are the {permittivities} of the media above and below the film; 
recall that it is assumed 
$ \varepsilon '_{ m } < 0 $ and $ \left| \varepsilon_{ m } \right| \gg 1 $. 
This equation can be obtained from {an} $L - C$ 
model { if} we consider 
the metal film as an inductance that shorts 
displacement currents of the size
 $ \left| k_{p} \right|^{-1} $. 
The mean free path of {the} SP is estimated from Eq.\,(\ref{kp}) as 
$ l_{p}\simeq 1/ \Im (k_{p}) \simeq
d \left|\varepsilon_{ m } \right| /
\left( {(\varepsilon_{d}+ \varepsilon_{e})} \delta_{ m } \right) \gg d $,
where $ \delta_{ m } = \varepsilon''_{ m } /
\left|\varepsilon_{ m } \right|\ll 1 $
is the loss tangent. For example, in {the} red and infrared 
spectral {ranges,} the {mean free path} of {the} SP {satisfies}
$ l_{p} > 10^{2} \, d $ 
in a thin silver film. 
Equation (\ref{kp}) holds when the thickness $d$ of the metal film
is less than the skin depth $ d < \delta_{sk} $, where
$\delta_{sk} \sim 1/ \left( k \sqrt{ - \varepsilon'_{m}} \right) \sim 
c/\omega_{p}$; the last {estimate} is made for a Drude metal, where
$\omega_{p} $ is the plasma frequency. 
For example, the skin depth {is estimated as} $ \delta_{sk} < 50\,nm $ 
for silver, gold, and copper films in optical and IR bands 
\cite{JohnsonChristy1972}.
In {a ``thick'' metal film whose} thickness $ d > \delta_{sk} $,
two weakly connected SPs propagate on both sides of the film.
Their wave vector $ k_{p} $ is slightly larger than
 {the free-space} wave vector. 

In a smooth metal film the incident light cannot excite {an} SP 
 since its speed $c_{p}= \omega / k_{p}$
is always less than the speed 
of light, and therefore phase matching is unattainable.
Unlike a flat surface, SP propagating in 
a metal shell has {radiative 
losses} and{, therefore,} it {can} be excited by the incident light. 
The local electromagnetic field 
is significantly enhanced when the shell SP is activated, 
as shown in Figs.\, \ref{Fig1},  \ref{Fig2}, and  \ref{Fig3}.
Substituting Eq.\,(\ref{dRes0}) {into} Eq.\,(\ref{kp}){,} we obtain 
the resonance condition 
\begin{equation}
	\label{kpa=q}
	 {k_{p} \, a \simeq
	 \frac{q(q+1)(\varepsilon_d+\varepsilon_e)}
	 {q\varepsilon_d+(q+1)\varepsilon_e}
	 \xrightarrow{\ q\gg1\ } q},
\end{equation}
where $ q = 1, 2, 3 \ldots $ {is an integer denoting} 
the order of the shell plasmon 
excited in the metal envelope. 
 {Equation (\ref{kpa=q}) approaches the scaling 
	$k_pa\sim q$ in the
large-$q$ local approximation. It could be interpreted as a
count of plasmon wavelengths around the circumference.}

When the envelope thickness $d$ is larger {than} the skin depth 
$d_{sk}${,} the plasmon wavevector $k_{p}$ 
is similar to the {free-space} 
wavevector $2 \pi / \lambda$. 
Core--shell systems then resonate when the wavelength of the incident
 light is close to the circumference of the metal shell. 
 Most FON substrates, starting with the seminal work 
 by Van Duyne et al.~\cite{VanDuyneHulteenTreichel1993}, consist of microspheres whose diameter is 
 {on} the order of the wavelength of the optical radiation.
 The spheres are covered by a thick metal film and 
 can be considered as an array of core--shell resonators.
We believe that collective shell plasmons have been 
observed in recent real-world and computational experiments
with gold and silver FONs \cite{LinLiaoChen2011,
GreeneltchBlaberHenry2013,HernandezReynosoFainstein2023, HernandezSalasParada2025,SusterSzymanskaAntosiewicz2025}.
The excitation of shell plasmons and the corresponding
electric field enhancement make FONs very efficient substrates for
 SERS. It should be noted that the quasi-periodic structure 
reduces radiative losses in FON substrates and increases 
the external local field.

 {A thin metal shell} enhances the field in the dielectric core.
We believe that resonances in the thin shell have been
 {observed} in the experiments 
 \cite{SilvaNygaDrachev2015,MannNascimentoDuplatKeul2017}.
The optical quality of {a} core--shell plasmon resonator 
is determined by {ohmic losses} in the metal envelope
and by {radiative losses} of the whole core--shell {system}. 
The ohmic {losses 
decrease as} the metal thickness {decreases,} 
while the radiation {losses
can} be decreased by increasing 
the diameter and {correspondingly 
increasing the order} of the plasmon resonance
(cf. Figs.\,1a,\,2a and Figs.\,1b,\,2b). 

The electric field in {the core--shell system} is enhanced in the internal volume
$V_{p} \sim \lambda_{p} a^{2} \sim a^{3}/q ${, as seen} in 
Figs.  \ref{Fig1} and  \ref{Fig2}. {That is,}
the volume $V_{p}$ is much 
larger than the volume of a gap plasmon in a random
SERS substrate, composed {of} metal nanoparticles (see, e.g.,
 \cite{SarychevBarbillonIvanov2023,
 StefancuAizpuruaAlessandri2024}
and references therein). 
 {The large} volume $ V_{p} $ of the field maximum 
 {makes it possible to achieve
field enhancement throughout the volume 
of a micro-object such as a 
cell, exosome, virion, or microplastic particle. 
This requires only that} 
the micro-object {be covered with a thin metal shell}.

In this paper we have calculated plasmon resonances, 
which are excited by {a} plane EM field. Therefore, 
all the fields are expanded in spherical harmonics with 
$ P_{q}^{1} (\cos \theta ) \cos \varphi $ 
and 
$ P_{q}^{1} (\cos \theta ) \sin \varphi$ functions 
of the polar angle $ \theta $ and the azimuthal angle 
$\varphi$ (see Eqs.\,(\ref{EleSpPot}) 
and (\ref{MagSpPot})). 
Dispersion equations, which are similar to Eqs.\,(\ref{DetE}) 
and \,(\ref{DetH}), can 
be obtained for the plasmon resonances with 
$ P_{q}^{m} (\cos \theta ) \cos m \varphi $
 and 
 $ P_{q}^{m} (\cos \theta ) \sin m \varphi $ 
 harmonics, where the order $ q $ is {an} arbitrary positive integer and 
 the azimuthal number $ 0 \le m \le q$ 
is {a} non-negative integer. 
Since the radial dependence of the Mie harmonics 
is independent of the azimuthal number $m$,
the determinants (\ref{DetE}) and (\ref{DetH}) 
of the equations for stitching fields in the core, shell, 
and surrounding space {vanish} simultaneously 
for all $ m$ {at a} given order $ q $.
The resonances with {an} azimuthal number $m$ other than {1}
cannot be excited by a simple plane 
electromagnetic wave. Sometimes they are called dark modes 
or {bound states in the continuum} 
\cite{StockmanKneippBozhevolnyi2018,BergmanStockman2003,
KoshelevBogdanovKivshar2020}.
For {example,} the electric and magnetic fields in the ``spherical''
$ q = 1, \, m=0 $ plasmon mode {are} similar to {the} EM field in {a} nonradiating
 {monopole} electric source \cite{ZanganehEvlyukhinMiroshnichenko2021}.
However, in a typical SERS experiment, a plasmonic resonator
is excited by {the EM field produced by a strongly focusing lens}. It also interacts 
with nearby resonators. Therefore, {a} core--shell resonator is 
typically located in a complex, nonuniform near-field. 
 {A dependence} of SERS on the numerical aperture of the focusing lens
 was observed, e.g., in~\cite{DuKasimYou2011}.
Accessible plasmonic resonances can be excited by the 
nonuniform field{, as discussed}, e.g., in
 \cite{ZuritaSanchez2019,ZuritaSanchez2021}. 
The total number $N$ of shell
plasmons can be estimated from the specific phase volume
$ N \sim (4 \pi a^{2})(\pi k_{p}^{2})/(2 \pi)^{2}
\sim (a k_{p})^{2} \sim q^{2}$, 
where the first brackets contain the surface area of
the metal shell over which {the plasmon propagates}. 
The density of states $\rho \sim \dfrac{d\,N}{d\,q} \sim 2 q$
 {is only a continuum estimate; the exact 
 	degeneracy of one spherical
multipole family is $g_q=2q+1$}.
For {example, the} hexapole resonance ($q = 3$) 
contains {seven}
different plasmons. Excitation of the "dark" plasmons results 
in {additional} electric field enhancement.

The use of {metal} shells over dielectric macromolecules such as 
protein globules offers several advantages 
for acquiring SERS signals. 
A {metal} shell absorbs and dissipates heat, protects proteins from 
laser-induced denaturation, and acts as an armour, shielding proteins 
from organic solvents. At the same time, the {metal} shell functions
 as a nanoantenna that enhances Raman scattering from the protein 
 molecules. Sarychev \textit{et al.} engineered an {ultra-narrow} 
 metal--dielectric nanocavity consisting of a silver-coated film 
 of the receptor-binding domain (RBD) of the 
 SARS-CoV-2 spike (S) glycoprotein and an 
 underlying silver surface, the two components 
 being held together by interactions between reduced 
 protein sulfhydryl groups and silver. 
 The authors demonstrated that the protein SERS 
 signal can be increased by several orders of 
 magnitude by coating the RBD film with a 
 nanometre-thick silver shell, thereby boosting 
 the cavity quality factor ($Q$-factor). 
 The strong light confinement in this nanocavity 
 enabled label-free acquisition of characteristic 
 Raman spectra from protein samples smaller than
 1~pg~\cite{KurochkinEremenkoEvtushenko2021,%
 SarychevSukhanovaIvanov2022}. 
 \subsection{SERS in {a single core--shell PMS}}
 \label{SERS in single core -- shell PMS}
\begin{figure}[!htbp]
 \centering
 \includegraphics[width=\textwidth]{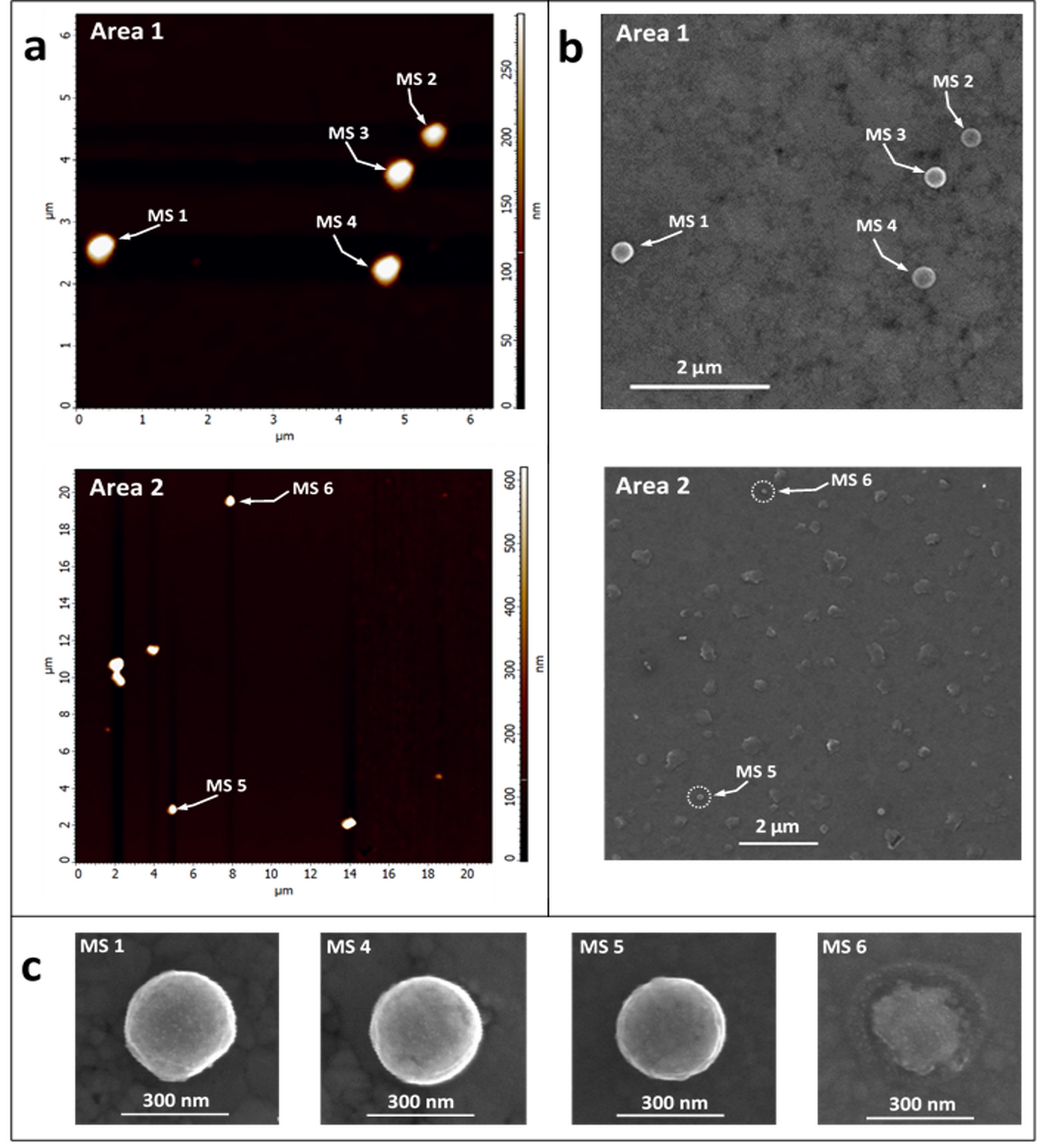}
 \caption{ 
 	AFM and SEM characterization of polymer-based 
 metal–dielectric resonators on a planar Ag film. Spectroscopy was 
 performed on six marked resonators (MS1–MS6).
 (a) AFM images acquired before deposition of the 
 $ 20\,nm$ Ag top layer: 
 Area 1 (MS1–MS4) and Area 2 (MS5, MS6).
 (b) SEM images of the same areas after deposition of the 
 $20\,nm$ Ag top layer.
 (c) High-magnification SEM images of MS1, MS4, MS5, 
 and MS6 (all are post-deposition), corresponding to the structures 
 compared in Figs.\, \ref{Fig6} and  \ref{Fig7}.}
 \label{Fig5}
\end{figure}
\begin{figure}[!htbp]
 \centering
 \includegraphics[width=\textwidth]{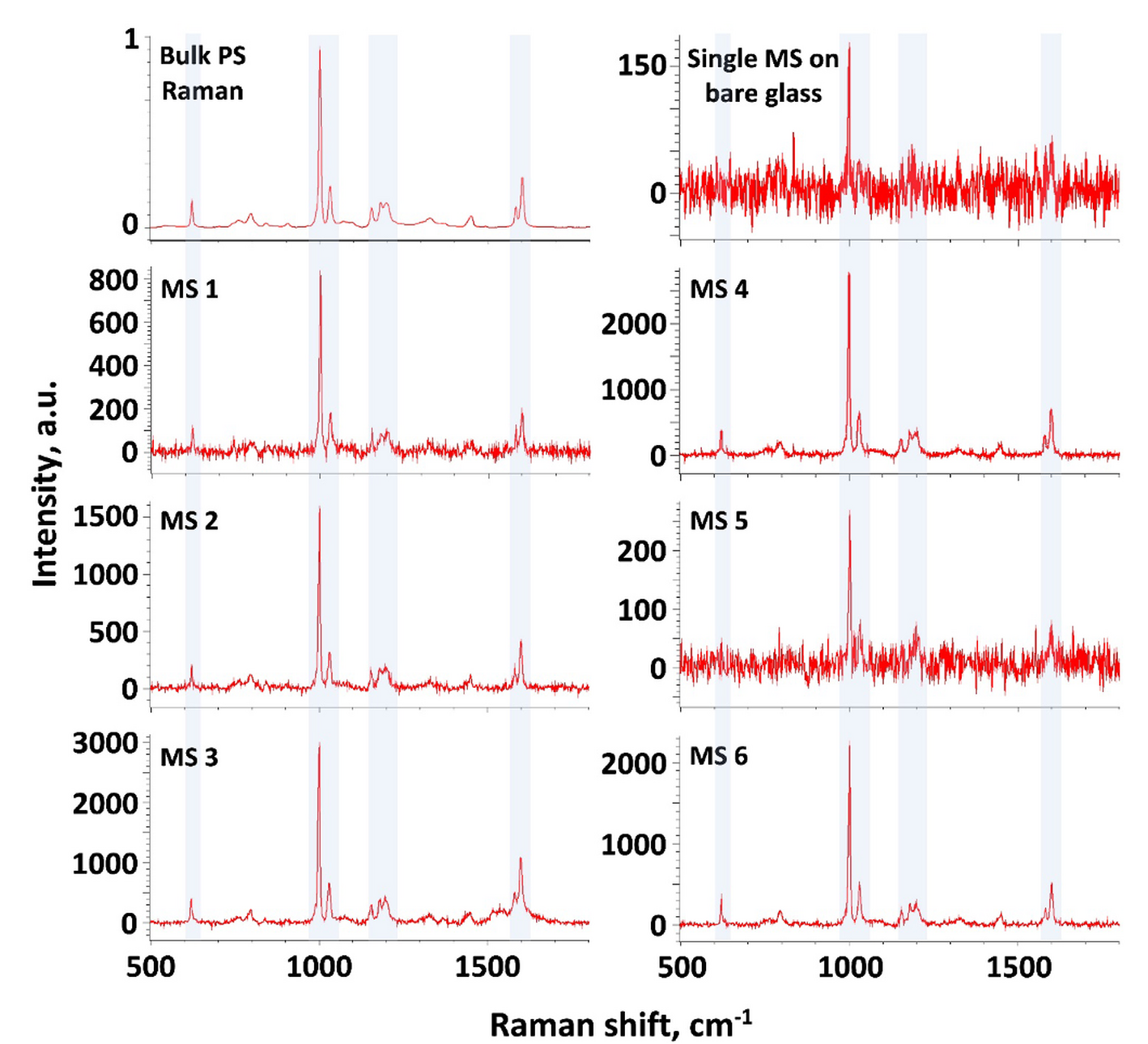}
 \caption{ 
 Raman and SERS spectra of PMS deposited 
 onto a flat silver layer prior to deposition of the upper 
 20~nm Ag overlayer. The upper left panel shows the bulk 
 PMS Raman spectrum (normalized to unity) obtained 
 with an excitation laser power of 20~mW. 
 The upper right panel displays the Raman spectrum of a single PMS 
 recorded with the same excitation power (20~mW). 
 The six lower panels (MS1--MS6) present SERS spectra 
 of individual PMS acquired at the positions indicated in 
 Fig.\, \ref{Fig5}a, all measured with 
 an excitation laser power of 
 $20\,mW$. The most prominent spectral regions 
 of the PMS spectrum are highlighted by grey bands.}
 \label{Fig6}
\end{figure}
\begin{figure}[!htbp]
 \centering
 \includegraphics[width=\textwidth]{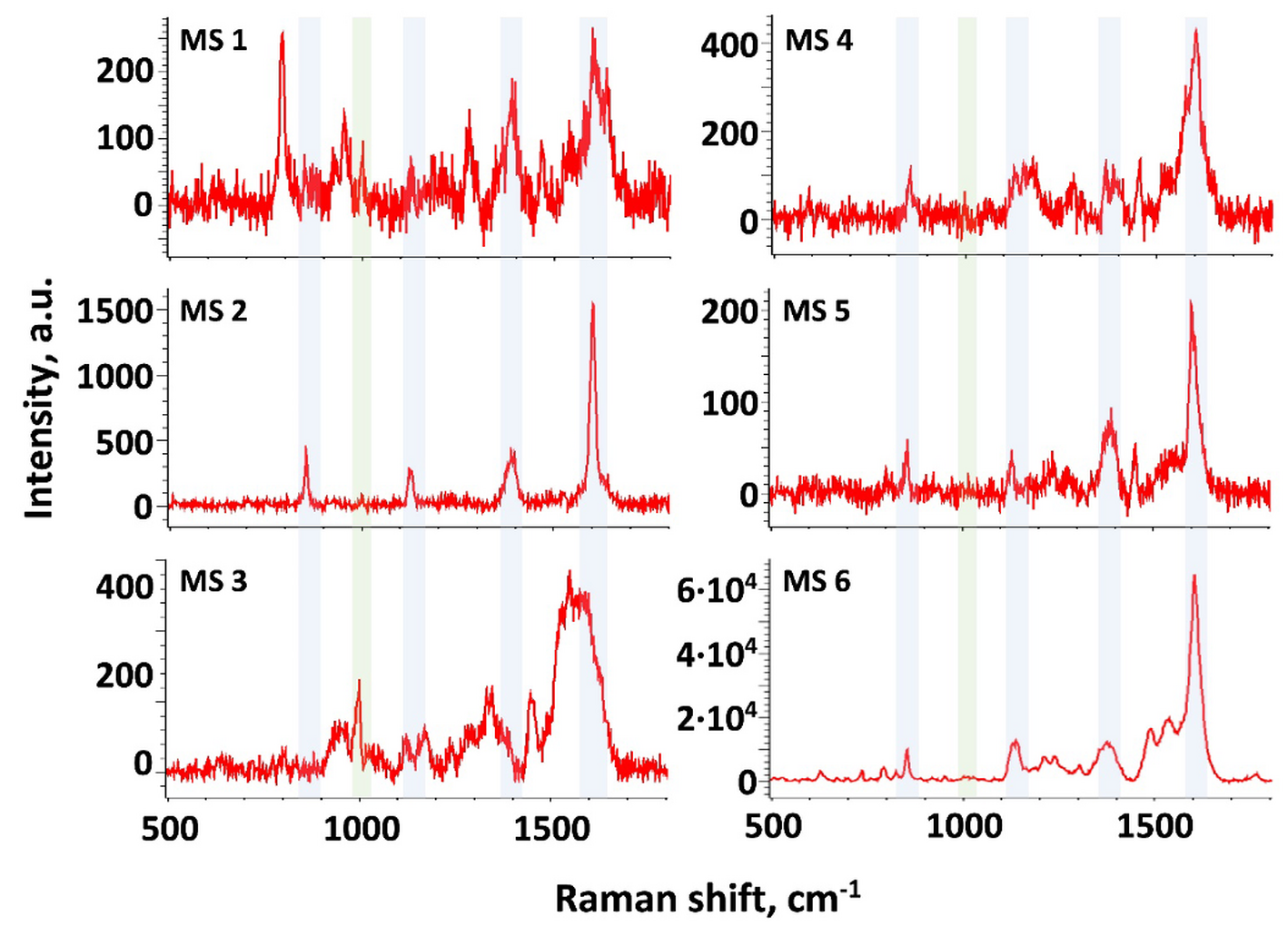}
\caption{
	SERS spectra of six individual PMSs after deposition of the upper
	$20\,\mathrm{nm}$ Ag layer. The spectra of MS1--MS6 were acquired
	at the positions indicated in Fig.~\ref{Fig5}(a) using an excitation
	laser power of $0.5\,\mathrm{mW}$. The most prominent spectral
	regions of polystyrene are highlighted by shaded bands.
}
 \label{Fig7}
\end{figure}
Here we perform a qualitative analysis of the interaction between 
an electromagnetic field at {a} wavelength of 532 nm and 
 {a} single PMS placed on {a} silver substrate. 
Two configurations are considered: a bare PMS 
on a flat silver surface and {a} PMS coated 
with a $20\,nm$ silver shell on the same substrate. 
A dielectric microsphere can act as a dielectric resonator by exciting 
Mie resonances. However, it is easy to verify that a sphere 
with a refractive index of 1.6 and a diameter of $300\,nm$ 
 {exhibits only weak optical resonances} when it is 
illuminated by a plane EM wave 
 \cite{BohrenHuffman2008}. Figures\,\ref{Fig4} a and b show
 the electric field distribution in the PMS ( $ 2 a = 300\,nm$, 
 $n_{1 } = 1.6 $, $\lambda = 532 \, nm$) located 
 on a smooth, optically opaque silver surface. It is shown 
that the electric field is localized {in} the center 
 of the sphere and on its side surfaces. The field is amplified
 approximately fourfold at normal incidence{,} and the electric field
increases approximately 20-fold when the light is incident at an angle of 50 degrees.
In contrast, when the sphere is coated with a continuous $20\,nm$ 
silver shell, plasmon resonances are excited, resulting 
in an enhancement factor of up to $ 10^{2} $ {(Fig.\,\ref{Fig4}b)}. 
The quality factor of the resonator increases in this case, 
with the maximum field observed along the inner shell surface.
Thus, it can be concluded that encapsulating 
a dielectric sphere with a continuous silver layer significantly 
enhances the local electric field in agreement with our theory.
The enhancement of the inner and outer electric {fields} is expected to 
increase RS from the core {material} as well as from molecules
deposited {on} the silver shell.

Figure\,\ref{Fig5} presents AFM and SEM 
images of individual PMS 
deposited onto a smooth $200\,nm$ silver surface. 
AFM data were acquired 
prior to deposition of the additional $20 \, nm$ 
top silver layer, while SEM 
images were recorded afterwards. This sequential acquisition was 
necessary to ensure that deposition of the silver overlayer {did} not 
introduce significant morphological changes to the sample, which is 
clearly demonstrated by the comparison of panels a and b in 
Fig.\,\ref{Fig5}. It can also be seen that the 
distance between individual 
PMS is on the order of one micron or more.
Taking into account the procedure for aligning the AFM 
and optical microspectroscopy fields of view, we can therefore 
state that{,} when recording a Raman spectrum from a single 
PMS{,} the contribution from neighboring spheres is minimized. 
Enlarged SEM images of four PMS are shown in 
Figure\,\ref{Fig5}c {to analyze} the dependence of SERS
 enhancement efficiency on the morphology of the upper 
 $20\,nm$ silver shell.

Figure\,\ref{Fig6} shows Raman scattering spectra from individual 
PMS deposited on the smooth silver surface. 
For comparison, the upper part of 
Figure\,\ref{Fig6} displays Raman spectra of bulk polystyrene 
and of a single polystyrene PMS. Importantly, all spectra presented 
in this figure are identical in terms of both peak positions and relative 
intensities. In particular, Figure\,\ref{Fig6} 
highlights the most pronounced bands: 
$621\,cm^{-1}$ corresponding to the benzene ring deformation mode,
 $1002\,cm^{-1}$ corresponding to the ring breathing mode,
 $1032\,cm^{-1}$ corresponding to C--H in-plane deformation, 
 a set of bands centered at $1180\,cm^{-1}$ corresponding to 
 C--C stretching, $1583\,cm^{-1}$ 
 corresponding to C=C stretching, and 
 $1602\,cm^{-1}$ corresponding to the ring skeletal stretch
 \cite{BridgesHoulneHarris2004}. Four of the six PMS demonstrate 
 Raman scattering enhancement factors up to 800 (MS3), 
 with the minimum enhancement around 150 (MS5), 
 compared to the spectrum of a single PMS on a bare glass 
 surface (upper right panel of Figure\,\ref{Fig6}). 
 In making this estimate, the {40-fold} difference 
 (20 vs. $0.5\,mW $) in excitation power between Raman 
 and SERS spectra was taken into account.
 
To evaluate the intrinsic SERS activity of the flat silver surface, 
we employed a standard analyte: a 0.5~mM aqueous stock solution 
of L-phenylalanine. A $30\,\mu\mathrm{L}$  droplet of this solution was 
deposited on the silver surface, and a specialized immersion
 micro-objective was focused onto the surface from above. 
 It was established that on truly smooth regions such as those
in Fig.\, \ref{Fig5}, the spectrum of L-phenylalanine 
 could not be detected, unlike in the case of defects such 
 as edges of holes in the silver film. 

 {A significant} enhancement of RS was reported in 
 SiO$_2$/TiO$_2$ and SiO$_2$/ZrO$_2 $ core--shell microresonators
 \cite{BontempiCarlettiDeAngelis2016,
BontempiVassaliniAlessandri2018}.
As discussed above{, self-assembled dielectric spheres enhance} RS 
on an array of metal nanoparticles
 \cite{YanXingJia2015,XingYanFeng2017,
HongZhouXu2020,Ogura2022,WangYanMi2022}.
All those studies considered 
micron-sized spheres that operate as {microlenses} 
\cite{WangGuoLi2011,ShinSeoLim2026}. 
 {As the diameter $2 a$ of a dielectric sphere increases,
	 higher-order EM
 resonances with $ q \sim n_{1} a/\lambda $
are excited.} 
As demonstrated above, electrodynamics simulations predict 
a modest electric field enhancement in the $ 2 a = 300\,nm$
PMS illuminated by {a plane wave incident normally on} the silver substrate
(see Fig.\, \ref{Fig4}a). 
We speculate that {the} local field can 
be further enhanced due to the
excitation of ``dark'' plasmon resonances {in} the {PMS--silver}
system. This suggestion is supported by simulation of the PMS when
the light is incident at an angle to the silver substrate, {as shown} 
in Fig.\, \ref{Fig4}(c). 
The Raman enhancement effect in PMS
placed on a silver mirror, demonstrated in this work, 
clearly warrants further investigation.

\begin{figure}[!htbp]
\centering
\includegraphics[width = 8.8 cm]{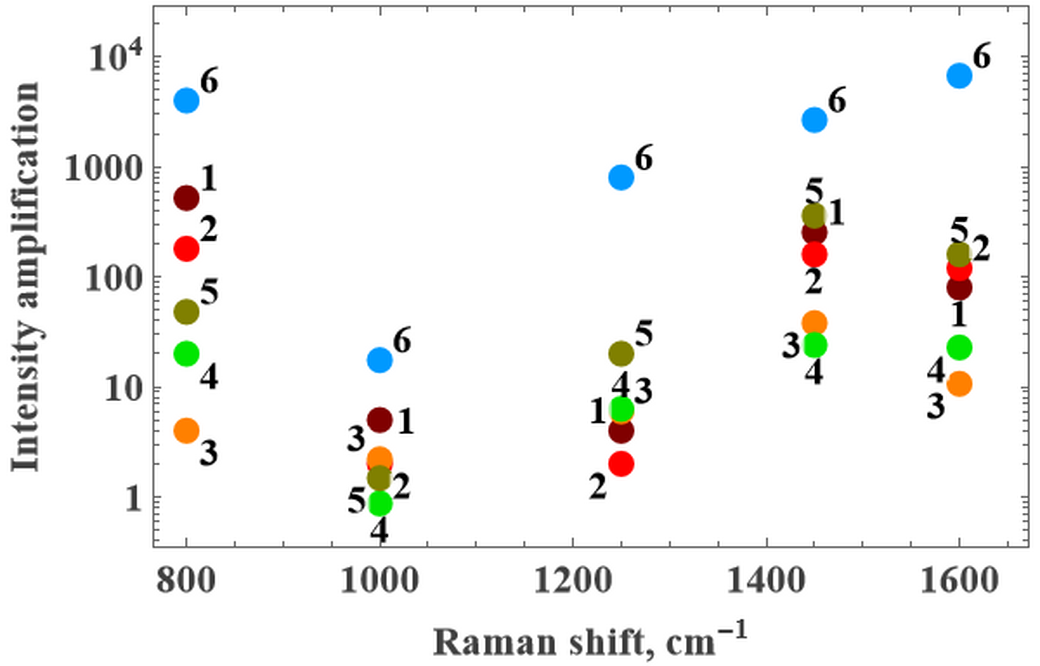}
	\caption{ 
Ratio of the amplitudes of Raman lines in PMSs 
covered with a $ 20\,nm $ 
silver shell to those of the corresponding lines in uncovered PMSs;
the PMSs are labeled as in Fig.\, \ref{Fig7}.}
\label{Fig8}
\end{figure}

The most interesting results are obtained when {a} $20\,nm$ silver 
shell is deposited {onto} {the PMS}.
Figure\,\ref{Fig7} shows SERS spectra of the same six 
PMS shown in Fig.\, \ref{Fig5}. 
Preliminary analysis of these spectra demonstrates 
a significant increase in SERS:
a Raman signal similar to that observed before silver deposition 
is now achieved at a forty-fold lower excitation power.
The Raman enhancement confirms the predictions 
of our theoretical analysis in 
Sec.\,\ref{Sec Results and Discussion}. 
{The SERS spectra are discussed in detail below.}
The silver shell enhances the Raman signal by several orders 
of magnitude. 
The ratio of the SERS from the PMS before and after deposition 
of $20 \,nm $ of silver shell is shown in Fig.\, \ref{Fig8}. 
The core--shell enhancement is uneven across the spectrum. 
SERS peaks at Raman {shifts} of $800\,cm^{-1}$ 
and $1600\,cm^{-1}$,
corresponding to wavelengths of $555\,nm$ and $ 580\,nm$,
are most prominent. {This suggests} that the silver shell provides
maximum field enhancement at these wavelengths.

Currently, we have not achieved complete 
 {reproducibility of SERS among} 
different core--shell PMSs. The maximum enhancement 
is achieved for MS6, shown in Fig.\, \ref{Fig5}.
Morphological analysis of core--shell PMS MS6 
in comparison with MS1, MS4, 
and MS5, which did not exhibit ultrahigh enhancement 
of {the} Raman signal, reveals differences that appeared 
after deposition of the silver shell and SERS acquisition. 
As shown in Fig.  \ref{Fig5}c, MS6 largely lost its initial spherical 
shape and appears slightly melted, while still retaining the 
upper {metal} coating. No morphological differences 
are observed between silver shells {deposited} on 
polystyrene spheres MS1, MS4, and MS5. 

It can be hypothesized that the enormous SERS enhancement, 
which is even greater than the maximum plasmon 
enhancement predicted by the Mie theory developed in Section
 \ref{Sec Mie expansion of EM field in core--shell microsphere}, 
is due to the formation of a graphitized polystyrene layer at the
interface between the silver shell and the polystyrene core. 
The inner surface of the metal shell can be uneven 
(see, e.g., Fig.\, \ref{Fig5}, MS6). 
The shell's roughness leads to coarse
graphitization of the polystyrene core when the metal shell is
heated by the laser beam. A fractal graphite structure can even form,
stabilized by heat transfer from the fractal to the
polystyrene core (see
\cite{YanShiJia2024}). {Despite substantial losses, a} 
surface plasmon can exist {at} the interface
between the graphite structure and the 
polystyrene \cite{Smolyaninov2022}.
We would like to note that a very rough metal surface 
is sometimes formed by the action of a pulsed laser beam 
on a thin metal film  
\cite{AshitkovInogamovZhakhovskii2012,
InogamovZhakhovskyPetrov2013}.
Under intense laser irradiation, the film surface can undergo explosive restructuring followed by resolidification into a fractal morphology.
Though the stabilization mechanism is not clear for
the case  of  CW laser used in our experiments.
 
The formation of 
 a graphene-like layer{, in which} electrons have high mobility, 
 leads to the appearance of virtual states and an increase in 
 the density of states involved in SERS.
 
 The exact nature of the {anomalously large} 
 SERS in the core--shell system
 observed in our experiments is currently unclear. 
 It can be assumed that a plasmon excited in the metal shell 
 is absorbed by the polystyrene molecule, and {an} electron is virtually
 transferred to the graphene layer, where it then relaxes 
 to the {molecular ground state}. The amplitude of this Raman transition 
 increases with the number of intermediate electron states. That is, the
 {Raman amplitude increases as the 
 area of the ``graphene'' film increases.} 
 The diversity of SERS in core--shell PMSs (see Fig.\, \ref{Fig8})
 can be explained by the different effective sizes of the 
 "graphene" layer at the interface between the polystyrene core 
 and the silver shell.
 \subsection{ {Laser-activated} SERS}
 \label{Laser activated SERS}
It is well known that molecular RS is a linear optical
phenomenon. Above, we assumed that RS from a core--shell structure
depends on the size of the graphitized layer at the core--shell interface.
Thermal graphitization depends on the laser beam intensity. 
RS then becomes a "nonlinear" phenomenon since its specific value
depends on the amplitude of the incident light. 
When molecular {oscillations interact strongly} 
with the same metal{,}
they therefore start to interact {with one another}. 
Then{,} mode competition
 {may occur, and some modes may be strengthened} 
at the expense of competitors. Finally, {collective oscillation} 
can be achieved. 

\begin{figure}[!htbp]
 	\centering
 	\includegraphics[width=0.8\textwidth]{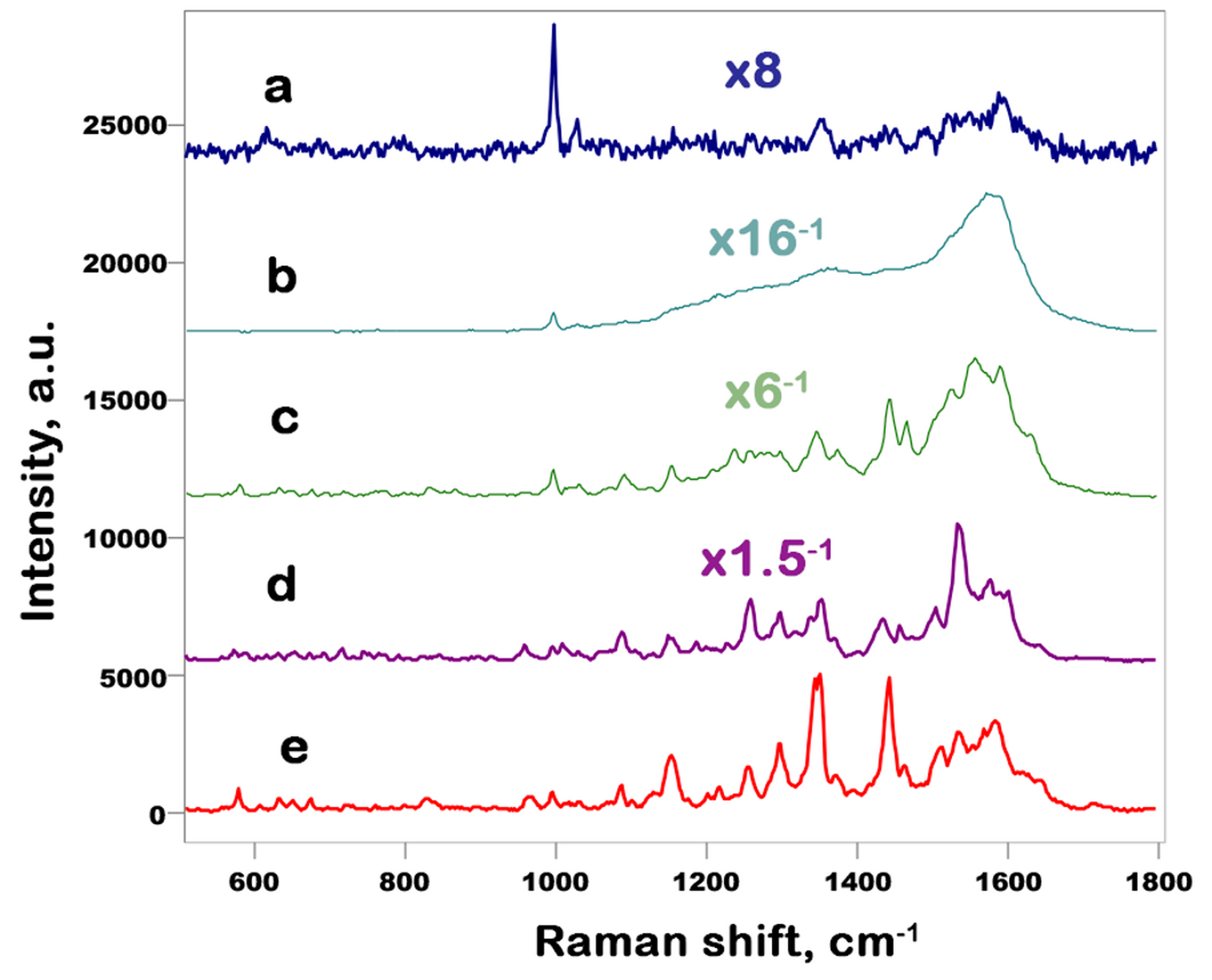}
 	\caption{ 
 		SERS spectra of a single ``heat-activated'' PMS 
 		deposited onto a flat silver layer after deposition of the upper 
 		$20 \, nm $ Ag overlayer. 
 		(a) SERS spectrum recorded before heat activation 
 		at an excitation laser power of $ 5\, m W $. 
 		(b) SERS spectrum recorded during the heat activation 
 		process at an excitation laser power of $ 20 \, mW$. 
 		(c--e) SERS spectra recorded after heat activation at excitation
 		 powers of 5, 1.25, and $0.5 \, m W$, respectively. 
 		For normalization, the intensities of the SERS spectra in panels
 		 (a--d) are multiplied by the factors indicated in each spectrum.}
 	\label{Fig9}
 \end{figure}

\begin{figure}[!htbp]
 	\centering
 	\includegraphics[width=0.8\textwidth]{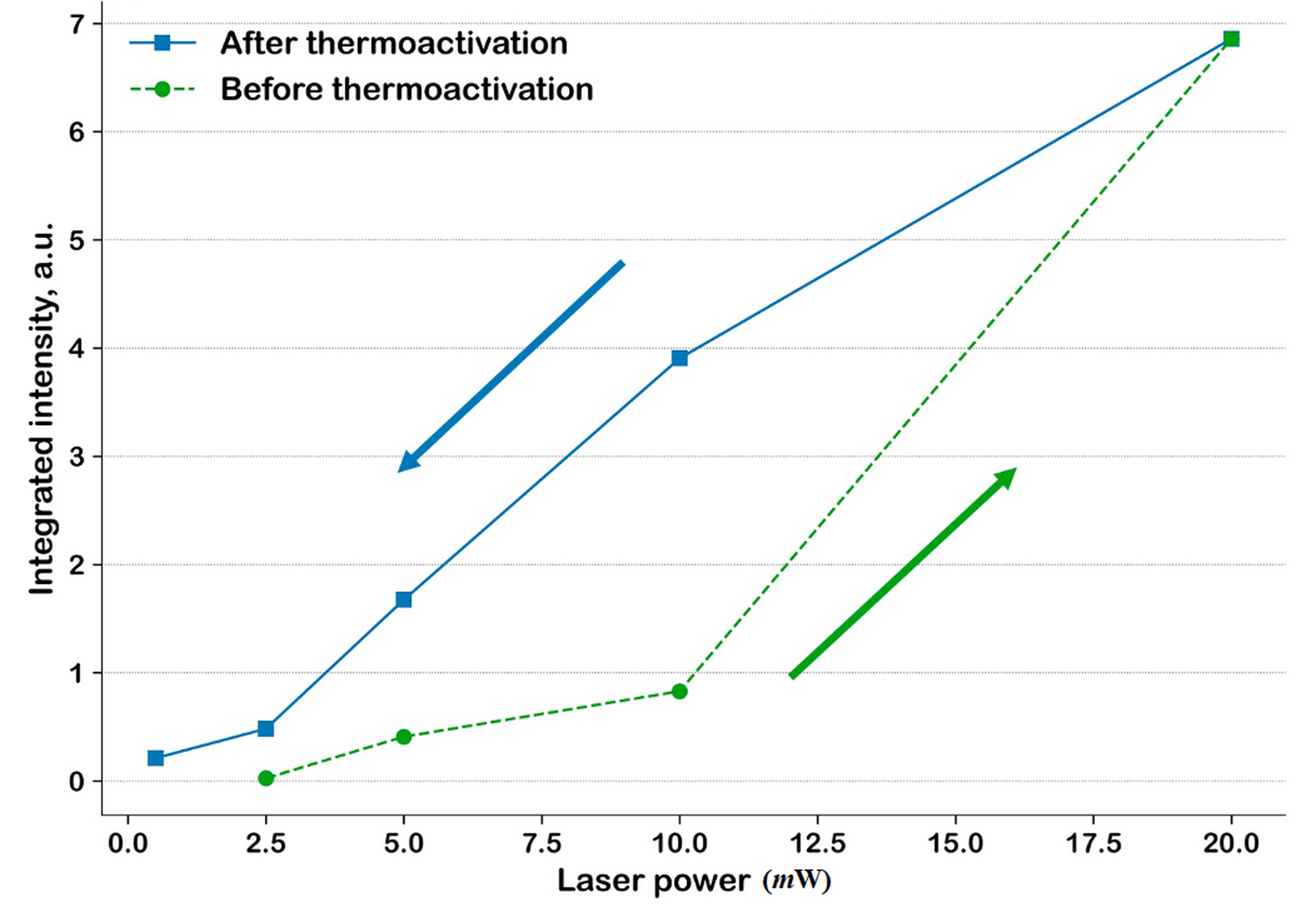}
 	\caption{ 
 		Dependence of the integrated SERS spectral intensity 
 		on the excitation laser power before (green) and 
 		after (blue) heat activation.}
 	\label{Fig10}
 \end{figure}

To test this hypothesis, we investigated how Raman scattering 
from an individual PMS depends on the laser power.
 {A} PMS that preserved the original polystyrene 
spectrum while exhibiting
enhanced intensity after deposition of the silver 
overlayer was investigated. We identified 
 {one} such microsphere {using} confocal microspectroscopy 
and monitoring the 
$1002\,cm^{-1}$ polystyrene marker band. 
Its spectrum (Fig.\,\ref{Fig9}a) 
indeed corresponds to that of polystyrene (Fig.\,\ref{Fig6}) 
and shows an intensity comparable to the spectra presented in 
Fig.\,\ref{Fig7}.
We next examined the effect of excitation laser power on this
 microsphere, with the results summarized in 
Figs.\,\ref{Fig9} and  \ref{Fig10}. In Fig.\,\ref{Fig9}a, 
the SERS spectrum is shown for the same excitation 
power as in Fig.\,\ref{Fig7}.
Spectra acquired at 2.5, 5, and $10~mW $ exhibit a linear 
dependence of integrated intensity on excitation power
(Fig.\,  \ref{Fig10}, green curve). At $ 20\, mW $, 
however, a pronounced nonlinear increase in integrated 
intensity is observed (Figs.\,  \ref{Fig9}b and  \ref{Fig10}). 
A twofold increase in the excitation power from $ 10\, mW $
to $ 20 \, m W $ results in a sevenfold enhancement of the SERS signal.
Acquisition at this power can be interpreted as 
a laser-activation process over 
25~s (1~s exposure, 25 accumulations). 
After a one-minute pause, spectra recorded at 10, 5, 1.25, 
and $ 0.5~m W $ again followed a linear power dependence
(Figs.\, \ref{Fig9}c–e and  \ref{Fig10}, blue curve).

We speculate {that,} upon laser activation, 
the {``PMS/metal-coating''} 
system undergoes a transition from a stationary inactive 
to a stationary active state. The spectra in this activated state
(Figs.\,\ref{Fig9}c–e) differ from those of polystyrene
(Fig.~\,\ref{Fig6}) and vary with excitation power. 
The effective SERS enhancement achieved 
through heat activation is comparable in both cases, 
approximately 40, suggesting a common mechanism that
probably depends {on} graphitization and {the} 
mode interaction discussed above.
 Nonlinear effects in SERS 
have been discussed, for example, in \cite{KneippKneipp2024,
 MuHwangChang2011,ShinLimShin2023,
 LombardiSchmidtWeller2018,RoelliGallandPiro2016,
 KharintsevKharitonovSaikin2017,LiuTieChen2022,
 KukushkinKirpichevMorozova2022,MaherCohenEtchegoin2004,
 HaslettTayMoskovits2000,YangLongWang2017,
 AndrianovVinogradovPukhov2025}. 
A similar effect of heating {PMSs} and changing 
their geometry under the influence of laser pump fields of different 
powers was observed in the work \cite{ShinLimShin2023}. 
The {authors} showed that nanoplastics can be locally melted 
on the SERS surface with increasing laser power, verifying 
the permeation of nanoplastics into a hotspot. Yet, the melting is
 {an} irreversible process{, whereas} 
 in our experiment the RS spectrum
is restored after {reducing} the laser power{, as clearly seen} 
in {Fig.\,\ref{Fig9}b--e}.

The results shown in Figs.\,  \ref{Fig9} and  \ref{Fig10} 
confirm the assumption made at the {beginning} of this subsection. 
Specific Raman scattering 
indeed increases with laser power. Moreover, the shape of the 
Raman spectrum also depends on {the} laser radiation, 
confirming our
assumptions about photomodification of the PMS. When the laser is 
turned off, the "graphene" layer degrades due to thermal fluctuations. 
Laser hysteresis is observed: as the laser power decreases, 
 {the Raman signal} 
also decreases, but remains higher than {that} 
in the core--shell before the thermal activation, 
as seen in Fig.\,  \ref{Fig10}.
 \FloatBarrier
 \section{Conclusions}
 \label{Sec Conclusions}
 In this work, we performed simulations 
 of {the} conditions required {to create} 
 an efficient system for the detection and spectral 
 analysis of globular and/or vesicle-like objects using SERS. 
 Interest in 
 developing such methodologies is driven by the need to establish
 analytical biomedical applications (for example, virus detection 
 and the identification of exosomal disease markers) as well as
 environmental monitoring methods, such as nanoplastic detection. 
 We modeled Raman scattering
 enhancement { using} PMSs with 
 a diameter of 300 nm as {model analytes}.
 In the first scenario, the analyte itself acted as part of the 
 SERS-enhancing system, namely as a dielectric spherical 
 microresonator interacting with a {metal} coating. It was established 
 that even {the} deposition of PMSs on a smooth silver surface, which 
 by itself does not provide plasmonic enhancement of Raman 
 scattering from adsorbed molecular analytes, leads to a
 two-order-of-magnitude enhancement of the Raman signal. 
 Importantly, the enhanced spectra obtained in this case are  identical to the unenhanced Raman spectra. 
 {The addition of a 20~nm-thick silver layer}
 in the form of spherical segments, designed to provide
 the highest possible gain in such structures, in most cases led
 to the predominance of the gain mechanism based on the excitation
 of plasmons in the silver shell.
 This approach provided {an additional two-order-of-magnitude 
 enhancement}. 
 
 We investigated the effect of laser activation of 
 {PMSs} by increasing the laser excitation power, 
 which {resulted} in an additional enhancement of the Raman 
 signal by nearly two orders of magnitude. Based on a detailed 
 analysis of AFM and SEM images, we speculate
 that laser activation could 
 involve {the creation
 of a graphene-like layer} at the silver-polystyrene interface. 
 
 Thus, the maximum enhancement obtained in this study is
 approximately six to seven orders of magnitude. 
 A four-order-of-magnitude enhancement of the original 
Raman spectrum of a PMS is achieved for the case 
of a metal–dielectric
 microresonator with a defect-free deposited silver overlayer.
 The results obtained indicate a clear direction for further 
 development: {creating} defect-free thin {metal} coatings 
 in close contact with the analyte and maximizing the contact area 
 between the metal and the dielectric globular analyte. 
 Examples of such work include the fabrication of profiled dielectric
 substrates (morphologically complementary to globular analytes) 
 coated with thin and homogeneous noble-metal films
 (\cite{LeeCarneyHazari2015,IvanovBykovBarbillon2024,
 	IvanovMochalovKorzhov2025}). 
 According to the results of our model experiments, such systems
 should provide Raman signal enhancement of globular analytes 
 on the order of {four orders of magnitude} 
 while preserving their original Raman spectra. We believe that the achieved performance level is sufficient for the further development of effective analytical methodologies.
\section*{Author Contributions}
\label{Sec Author contributions}
Andrey K. Sarychev: Conceptualization ({model} of the field
enhancement), Writing – review \& editing; 
Milena S. Shestopalova: Investigation (Sample preparation,
 {Raman \& SERS}), Formal analysis ({spectroscopy} data), 
Writing – review \& editing; Denis S. Korzhov: Investigation (AFM),
Data curation, Visualization; Andrey V. Ivanov: Formal analysis{,} Modeling,
Writing – review \& editing; Igor V. Bykov: 
Conceptualization ({the} idea of the upper silver layer {deposition}),
Writing – review \& editing; Konstantin N. Afanasev: 
Methodology (Silver layer deposition); 
Dmitry V. Basmanov: Resources, Writing – review \& editing; 
Nikita N. Osipov: Investigation (SEM); Vladimir A. Oleinikov: 
Writing – review \& editing; Konstantin E. Mochalov: Conceptualization, Supervision, Methodology, 
Writing – original draft.
\section*{Conflicts of Interest}
\label{Sec Conflicts of interest}
There are no conflicts to declare.

\section*{Data availability}
All data supporting the findings of this study are
	available within the article.

\section*{Acknowledgment}
\label{Sec Acknowledgements}
The authors are grateful to Professor A. N. Generalova 
for supplying polystyrene microspheres.



\balance


\bibliographystyle{rsc}
\bibliography{Allx15x09x2026}
\end{document}